\documentclass[11pt,showkeys,groupedaddress,nofootinbib,notitlepage,floatfix]{revtex4}
\usepackage{dsfont}
\usepackage{amssymb}
\usepackage{amsfonts}
\usepackage{amsmath}
\usepackage{graphicx}
\usepackage{subfigure}
\usepackage[dvipdfm,
            pdfstartview=FitH,
            bookmarksnumbered=true,
            bookmarksopen=true,
            colorlinks,
            pdfborder=001,
            linkcolor=green,
            anchorcolor=green,
            citecolor=red
            ]{hyperref}
\usepackage[toc,page,title,titletoc,header]{appendix}
\begin{document}
\title{Lifshitz Scaling Effects on Holographic Superconductors}
\author{Jun-Wang Lu$^{1}$}
\author{Ya-Bo Wu$^{1,2}$}
\thanks{E-mail address:ybwu61@163.com}
\author{Peng Qian$^{2}$}
\author{Yue-Yue Zhao$^{1}$}
\author{Xue Zhang$^{1}$}
\author{Nan Zhang$^{1}$}
\affiliation{
$^{1}$Department of Physics, Liaoning Normal University, Dalian, 116029, China\\
$^{2}$State Key Laboratory of Theoretical Physics, Institute of Theoretical Physics, Chinese Academy of Sciences, Beijing 100190, China}
\begin{abstract}
Via numerical and analytical methods, the effects of the Lifshitz dynamical exponent $z$ on the holographic superconductor models are studied in some detail, including $s$-wave and $p$-wave models. Working in the probe limit, we calculate the condensation and conductivity in both Lifshitz black hole and soliton backgrounds with a general $z$. For both the $s$-wave and $p$-wave models in the black hole backgrounds, as $z$ increases, the phase transition  becomes difficult and the conductivity is suppressed. For the Lifshitz  soliton background, when $z$ increases, the critical chemical potential increases in both the $s$-wave model (with a fixed mass of the scalar field) and $p$-wave model. For the $p$-wave model in both the Lifshitz black hole and soliton backgrounds, the anisotropy between the AC conductivity in different spatial directions is  suppressed when $z$ increases. In all cases, we find that the critical exponent of the condensation is always $1/2$, independent of $z$ and spacetime dimension.  The analytical results from the Sturm-Liouville variational method uphold the numerical calculations. The implications of these results are discussed.
\end{abstract}

\keywords{AdS/CFT correspondence, Holographic superconductor, Lifshitz gravity}
\maketitle

\section{Introduction}

The gauge/gravity correspondence provides us a powerful tool to study  the strongly coupled field theory via its dual gravity description~\cite{Maldacena1998,Gubser105,witten253}. Over the past years the gauge/gravity duality has been intensively used to study many systems in condensed matter physics.  One of interesting applications of the duality is to study high temperature superconductors, which are supposed to be
a strongly coupled system. The holographic $s$-wave superconductor model was first realized via an Einstein-Maxwell theory coupled to a complex scalar field in a  Schwarzschild-AdS black hole background~\cite{Hartnoll2008,Horowitz126008,Horowitz:2010gk,Horowitz:2009ij}. The condensation of the scalar breaks the U(1) symmetry of the system, mimicking the conductor/superconductor phase transition. The authors of Ref.~\cite{Siopsis} analytically studied the superconductor phase transition near the critical point.
 By an SU(2) gauge field in the bulk, a holographic $p$-wave superconductor model was constructed in Ref.~\cite{Gubser2008a}, in which the condensed vector field breaks the U(1) symmetry (one of subgroup of SU(2)) as well as spatial rotational symmetry spontaneously. An alternative  holographic $p$-wave superconductor model is realized by condensation of a 2-form field in a five-dimensional gauged supergravity~\cite{ARR}. The effect of the Gauss-Bonnet term on the $p$-wave model is discussed in Refs.~\cite{RGC1,RGC2}. Very recently a holographic $p$-wave superconductor model has been constructed in an Einstein-Maxwell-complex vector field theory
 with a negative cosmological constant~\cite{Cai:2013pda,Cai:2013aca}, where a rich phase structure is found than the SU(2) model. In addition, the holographic $d$-wave superconductor models are also built by introducing a charged massive spin-two field propagating in the bulk~\cite{Chen:2010mk,Benini:2010pr,Kim:2013oba}.

On the other hand, the holographic insulator/superconductor phase transition was studied in a five-dimensional AdS soliton background coupled to a Maxwell field and a charged scalar field~\cite{Nishioka131}. It was shown that when the chemical potential is beyond a critical value $\mu_c$, the pure AdS soliton solution modeling the insulator with a mass gap becomes unstable and results in a new hairy soliton solution dual to a superconducting phase in the boundary field theory. Further studies based on this model can be found, for example, in Refs.~\cite{Akhavan086003, RGCai046001, RGCai126007, QYPan088, HFLi135, RGCai4828,Lee2092}. Here we stress that those studies are all
based on the AdS soliton background. However, for the $p$-wave model in the AdS soliton background, up to now,  the conductivity has been calculated only in the direction perpendicular to the condensed vector~\cite{Akhavan086003}. To see the anisotropy of the $p$-wave superconductor model, it is helpful to calculate the conductivity along the condensed vector.

Recently, the phase transitions in many condensed matter systems are found to be governed  by the so-called Lifshitz fixed points which exhibit the anisotropic scaling of spacetime
$t\rightarrow b^z t, \vec{x}\rightarrow b \vec{x}$ ($z\neq 1$), where $z$ is the dynamical critical exponent representing the anisotropy of the spacetime. The gravity description dual to this scaling in the $D=d+2$ dimensional spacetime was proposed in Ref.~\cite{Kachru}
\begin{equation}\label{asymlifshitzst}
ds^2=L^2\left(-r^{2z}dt^2+r^2d\vec{x}^2+\frac{dr^2}{r^2}\right),
\end{equation}
where $d\vec{x}^2=dx^2_1+\cdots+dx^2_d$ and $r\in(0,\infty)$. This geometry reduces to the AdS spacetime when $z=1$, while it is a gravity dual with the Lifshitz scaling as $z>1$.
The Lifshitz spacetime (\ref{asymlifshitzst}) can be realized by a massless scalar field coupled to an Abelian gauge field in the following action~\cite{Tylor}
\begin{equation}\label{Lifshitz action}
S=\frac{1}{16\pi G_{d+2}}\int d^{d+2} x\sqrt{-g}\left(R-2\Lambda-\frac{1}{2}\partial_\mu\varphi\partial^\mu\varphi-\frac{1}{4}e^
{b\varphi}\mathcal{F}_{\mu\nu}\mathcal{F}^{\mu\nu}\right).
\end{equation}
The generalization of (\ref{asymlifshitzst}) to the case with finite temperature is~\cite{Pdw09052678}
\begin{equation}\label{Lifshitz metric}
ds^2=L^2\left(-r^{2z}f(r)dt^2+\frac{dr^2}{r^2f(r)}+r^2\sum_{i=1}^d dx_i^2\right),
\end{equation}
where
\begin{eqnarray}
 f(r)&=&1-\frac{r_+^{z+d}}{r^{z+d}},\ \  \Lambda=-\frac{(z+d-1)(z+d)}{2L^2},\label{metric function}\\
 \mathcal{F}_{rt}&=&\sqrt{2L^2(z-1)(z+d)}r^{z+d-1},\ \ e^{b\varphi}=r^{-2d},\ \ b^2=\frac{2d}{z-1}.
\end{eqnarray}
The Hawking temperature of the  Lifshitz black hole is
\begin{equation}\label{HawkingT}
T=\frac{(z+d)r_+^z}{4\pi},
\end{equation}
where $r_+$ denotes the black hole horizon.  It is interesting to construct holographic superconductor model by using Lifshitz black hole solutions and to see the effect of the dynamical critical exponent on the properties of holographic superconductors. Indeed some works have been carried out on this topic, see, for example, Refs.~\cite{Brynjolfsson065401,Sin4617,Buyanyan,Abdalla1460,ZYFan2000,RGCai066003,Hartnoll:2012pp}. In Ref.~\cite{Brynjolfsson065401} the authors simply studied the scalar condensation in a (3+1)-dimensional Lifshitz black hole background with $z=3/2$, while Ref.~\cite{Sin4617} constructed a $s$-wave superconductor model in a (3+1)-dimensional Lifshitz black hole spacetime with $z=2$. Bu in Ref.~\cite{Buyanyan} studied $s$-wave and $p$-wave superconductor models in the (3+1)-dimensional Lifshitz black hole spacetime (\ref{Lifshitz metric}) with $z=2$. Ref.~\cite{ZYFan2000} studied a $s$-wave model in
(3+1)-dimensional hyperscaling violation spacetime with $\theta=1$ and $z=2$. Recently,  Abdalla et al in Ref.~\cite{Abdalla1460} have investigated the $s$-wave superconductor phase transition in a three-dimensional Lifshitz black hole in new massive gravity with $z=3$ and found a series of peaks in the conductivity for certain values of the frequency.

 In this work, we are going to study systematically the effects of the Lifshitz dynamical exponent $z$ on the holographic superconductors based on the  Lifshitz spacetime (\ref{Lifshitz metric}) in the probe limit. The holographic models include $s$-wave and $p$-wave cases.  For all cases, there exists a critical temperature $T_c$, which decreases when $z$ increases ($z=1,~2$ in $D=4$ and $z=1,~2,~3$ in $D=5$). This indicates that the increasing $z$ inhibits the superconducting condensation. For the $p$-wave case, the difference between the AC conductivity in $y$ direction $\sigma_{yy}$ and in $x$ direction $\sigma_{xx}$ is suppressed as $z$ increases. We will also study the holographic insulator/superconductor phase transition
 in the Lifshitz soliton background, which is obtained by double Wick rotation to the  Lifshitz black hole spacetime (\ref{Lifshitz metric}). This part is totally new, there does not exist any relevant study in the literature.
  As $z$ increases, the insulator/superconductor phase transition becomes easy (the critical chemical potential decreases) in the $s$-wave model (with a fixed operator dimension) but difficult (the critical chemical potential increases) in the $p$-wave model.  In addition, we will study these superconducting phase transition by Sturm-Liouville variational method. The analytical method supports the numerical calculations.

This paper is organized as follows. In section II, we study the $s$-wave conductor/superconductor and insulator/superconductor phase transitions in the Lifshitz black hole and soliton backgrounds, respectively, by  numerical and analytical method. The $p$-wave conductor/superconductor (insulator/superconductor) phase transitions will be studied in Sec. III. The final section is devoted to the conclusions and  discussions.

\section{Holographic $s$-wave superconductors in the Lifshitz spacetime}

In this section, we first study the holographic $s$-wave superconductor model in the Lifshitz black hole background. To complement the numerical calculations, we also study the conductor/superconductor phase transition by the Sturm-Liouville variational method. In the second part of this section, we will study the holographic $s$-wave superconductor model in the Lifshitz soliton background.

Following Ref.~\cite{Hartnoll2008}, we consider the Lagrangian density consisting of a Maxwell field and a complex scalar field
\begin{equation}\label{Swave action}
  \mathcal{L}_m=-\frac{1}{4}F_{\mu\nu}F^{\mu\nu}-|D_\mu\psi|^2-m^2 |\psi|^2,
\end{equation}
where $D_\mu=\nabla_\mu-iq A_\mu$, $F_{\mu\nu}=\nabla_\mu A_\nu-\nabla_\nu A_\mu$ and $m$ ($q$) is the mass (charge) of the scalar field $\psi$.  From (\ref{Swave action}) we have the equations of motion of $\psi$ and the Maxwell field
\begin{eqnarray}
  D_\mu D^\mu\psi-m^2\psi&=&0,\label{EOMofpsiofSwave}\\
  \nabla^\mu F_{\mu\nu}-iq(\psi^\ast D_\nu\psi-\psi {D_\nu}^\ast\psi^\ast)&=&0.\label{EOMofphiofSwave}
\end{eqnarray}
We will work in the so-called probe approximation, namely the backreaction of the matter sector (\ref{Swave action}) on the background
Lifshitz geometry is neglected. In addition, by using the gauge symmetry in (\ref{Swave action}), we can consider the following
ansatz for the scalar field and Maxwell field as
\begin{equation}
\psi=\psi(r),  \ \ \  A_\mu dx^\mu=\phi(r) dt.
\end{equation}

\subsection{ $s$-wave superconductors in the Lifshitz black hole background}

In this subsection we study the holographic $s$-wave  superconductor model in the Lifshitz black hole background (\ref{Lifshitz metric}). In this case, the equations of motion (\ref{EOMofpsiofSwave}) and (\ref{EOMofphiofSwave}) in the background (\ref{Lifshitz metric}) reduce to
\begin{eqnarray}
 \psi''+\left(\frac{d+z+1}{r}+\frac{f'}{f}\right)\psi'+\frac{q^2\phi^2}{r^{2z+2}f^2}\psi-
 \frac{m^2L^2}{r^2f}\psi&=&0, \label{numbhspsi0inf} \\
 \phi''+\frac{d-z+1}{r}\phi'-\frac{2q^2L^2\psi^2}{r^2f}\phi &=& 0,\label{numbhsphi0inf}
\end{eqnarray}
where a prime stands for the derivative with respect to $r$. In the remainder of this paper, we will set $L=1$ and $q=1$. To solve the above equations, we have to specify the boundary conditions for the two fields. At the horizon $r=r_+$, we impose $\phi(r_+)=0$ to satisfy the finite norm of $A_\mu$, while $\psi(r_+)$ needs to be regular.
At the boundary $r\rightarrow \infty$, $\psi(r)$ and $\phi(r)$ behave as
\begin{eqnarray}
\psi(r)&=&\frac{\psi_1}{r^{\Delta_-}}+\frac{\psi_2}{r^{\Delta_+}}+\cdots, \label{asySwaveBHSpsi}\\
\phi(r)&=&\mu-\frac{\rho}{ r^{d-z}}+\cdots(z<d), \  \text{and}\  \  \mu-\rho\ln{\xi r}+\cdots (z=d), \label{asySwaveBHphi}
\end{eqnarray}
 where $\Delta_\pm =\frac{z+d\pm\sqrt{(z+d)^2+4m^2}}{2}$, $\xi$, $\psi_1$, $\psi_2$, $\mu$ and $\rho$ are all  constants. According to the gauge/gravity duality, $\psi_1$ ($\psi_2$) can be regarded as the source (the vacuum expectation value) of the dual operator $\mathcal{O}$, and $\mu$ and $\rho$ are chemical potential and charge density of dual field theory, respectively. Since we require that the U(1) symmetry is broken spontaneously, we impose the source-free condition $\psi_1=0$. We denote $\Delta=\Delta_+$ throughout the paper. The mass squared $m^2$ of the scalar field has a lower bound as $m^2=-(z+d)^2/4$ with ${\Delta}={\Delta}_{BF}=(z+d)/2$. In that case, there is a logarithmic term in the asymptotical expansion (\ref{asySwaveBHSpsi}). We treat the coefficient of this logarithmic term as the source which is set to be zero to avoid the instability induced by this term following Ref.~\cite{Horowitz126008}.

 In this paper we consider canonical ensemble where $\rho$ is fixed, when we discuss the black hole backgrounds. Concretely we focus our numerical calculation on the cases of $z=1,~2$ in $D=4$ and $z=1,~2,~3$ in $D=5$. To see clearly the effect of the dynamical critical exponent $z$, we fix the dimension $\Delta$ of the boundary scalar operator.
\begin{figure}
\begin{minipage}[!htb]{0.45\linewidth}
\centering
\includegraphics[width=2.8in]{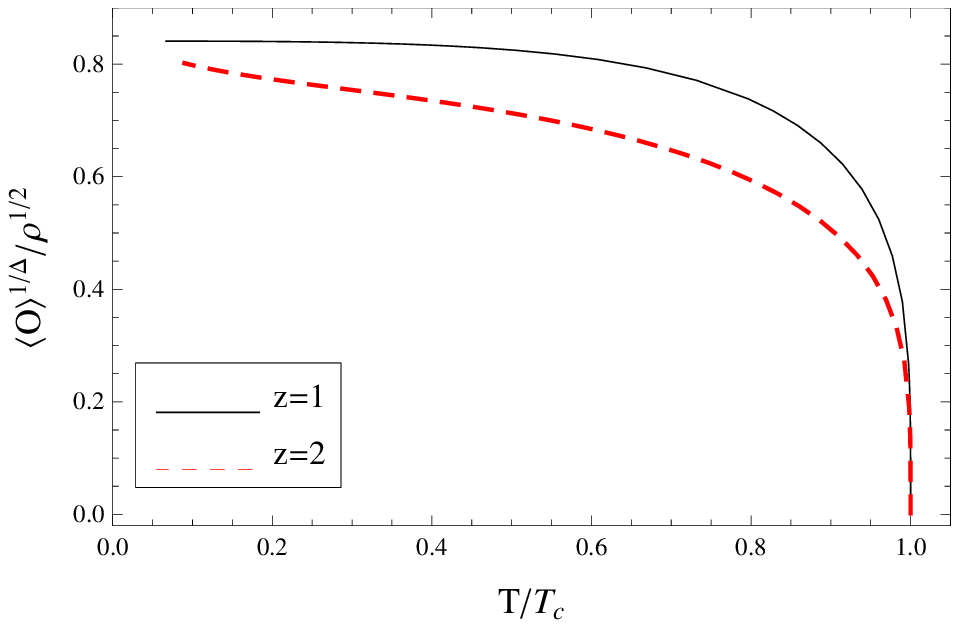}
\end{minipage}
\begin{minipage}[!htb]{0.45\linewidth}
\centering
 \includegraphics[width=2.8 in]{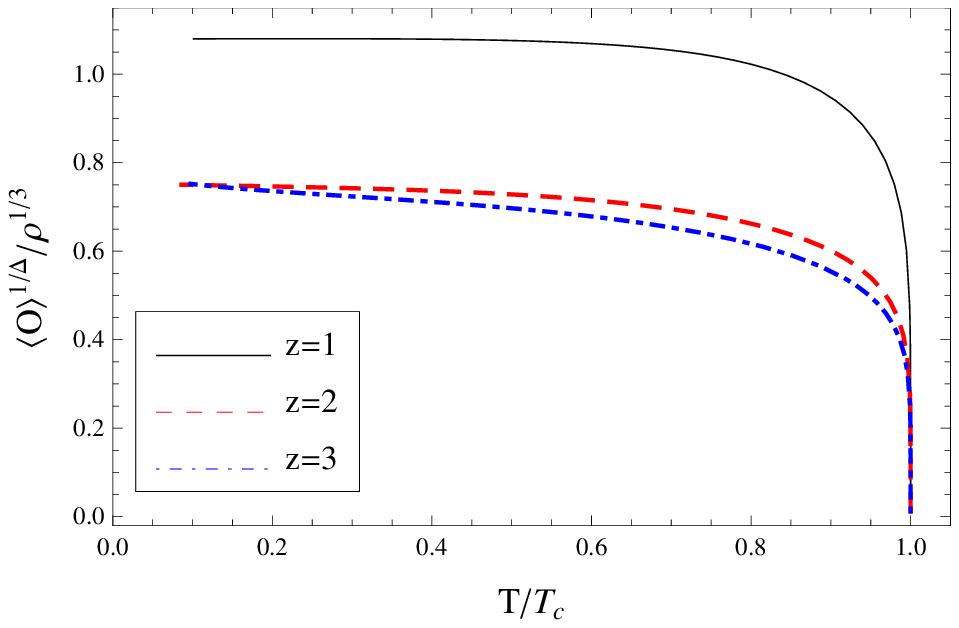}\\
\end{minipage}
\caption{The condensation  versus temperature in the  $s$-wave model for $\Delta=2$ in $D=4$ (left) and $\Delta=3$ in $D=5$ (right). The curves from top to bottom in the left plot correspond to $z=1$~(solid),~2~(dashed), while the ones in the right plot to $z=1$~(solid),~2~(dashed),~3~(dotdashed), respectively.}
\label{dgzgbhSwavecsct}
\end{figure}
Figure~\ref{dgzgbhSwavecsct} shows the condensation as a function of temperature for various $z$, from which we can see that the condensation decreases with the increase of $z$.  Note that in the $D=5$ case, the curves of the condensation for $z=2$ and $z=3$ intersects at some low temperature. In fact in the numerical calculations, we find that the condensation for the case of $z=2$ in $D=4$ and of $z=3$ in $D=5$  increase slight quickly at low temperature than other curves. This might be due to the fact that for the cases $z=2$ in $D=4$ and $z=3$ in $D=5$, there is a logarithmic term in the gauge field $\phi$ expansion near the boundary $r \to \infty$. In addition, at the sufficiently low temperature, the backreaction effect of the matter sector on the background geometry becomes important, thus the probe approximation considered in this paper is no longer valid. For a comparison,  we list in Tab.~\ref{tab:BHswavenumericalresults} the critical temperature $T_c$ and  the condensation behavior near $T_c$ for the cases of $z=1$, $2$ and $3$ with different operator dimension.
 \begin{table}
\caption{The critical temperature, condensation and superfluid density  for the $s$-wave superconductor in the 4(5)-dimensional Lifshitz black hole backgrounds.  Here $t=1-T/T_c $,  the subscript $ {SL}$ denotes the quantity calculated by the Sturm-Liouville method, and $\langle\mathcal{O}\rangle^{1/\Delta}/\rho^{1/d}$ and $\tilde{n}_s=n_s/\rho^{(d+z-2)/d}$ as well as $\langle\mathcal{O}\rangle^{1/\Delta}_{SL}/\rho^{1/d}$ are calculated near $T_c$.}
\label{tab:BHswavenumericalresults}
\begin{ruledtabular}
\begin{tabular}{c  c c  c c c  c c}
  $D$& z & $m^2$ &$T_c/\rho^{z/d}$& $\langle\mathcal{O}\rangle^{1/\Delta}/\rho^{1/d}$ &$\tilde{n}_s$ & $T_{c;SL}/\rho^{z/d}$&$\langle\mathcal{O}\rangle^{1/\Delta}_{SL}/\rho^{1/d}$  \\ \hline
     4 &1& $-2$  &$0.118$ & $1.19 t^{1/2\Delta}$ & $2.82 t$& 0.117&0.95$t^{1/2\Delta}$ \\
     4&2& $-4$ &$0.068$ & $ 0.92t^{1/2\Delta}$ & $1.95 t$ & ---&--- \\
        4 &2& $-3$ &$0.035$ & $ 0.66t^{1/2\Delta}$ & $0.88 t$ & ---&--- \\
     5&1& $-15/4$  &$0.220$ & $ 1.56t^{1/2\Delta}$& $5.19t$  &0.218&$1.36t^{1/2\Delta}$  \\
     5&1& $-3$ &$0.197$ & $ 1.44t^{1/2\Delta}$ & $4.49t$ &0.196&$1.22t^{1/2\Delta}$  \\
     5&2& $-6$  &$0.087$ & $ 0.90t^{1/2\Delta}$& $1.53 t$  &0.087&0.94$t^{1/2\Delta}$  \\
     5&3& $-9$ &$0.045$ & $ 0.82t^{1/2\Delta}$ & $0.93 t$ & ---&--- \\
\end{tabular}
\end{ruledtabular}
\end{table}
From the table, we can find that when we increase $z$, $T_c$ decreases for the case with a fixed $\Delta$, which indicates that the increasing anisotropy between space and time hinders the phase transition. This can be understood as follows.
 We can see from Eq.~(\ref{numbhspsi0inf}) that near the horizon, the effective mass of the scalar field increases as the dynamical critical exponent $z$ increases. This leads to a lower critical temperature as $z$ increases.  Here we mention that the condensation in the case with $z=2,~m^2=-3,~D=4$ is also calculated in Ref.~\cite{Sin4617}, but the latter works in a grand canonical ensemble, while in the cases with $z=2,~m^2=-3~(0),~D=4$, our results are consistent with the ones in Ref.~\cite{Buyanyan}. On the other hand, we can see from Tab.~\ref{tab:BHswavenumericalresults} that all curves of condensation versus temperature have a square root behavior near $T_c$, which suggests that the critical exponent is $1/2$,  as expected from the mean field theory.

\begin{figure}
\begin{minipage}[!htb]{0.43\linewidth}
\centering
\includegraphics[width=2.75in]{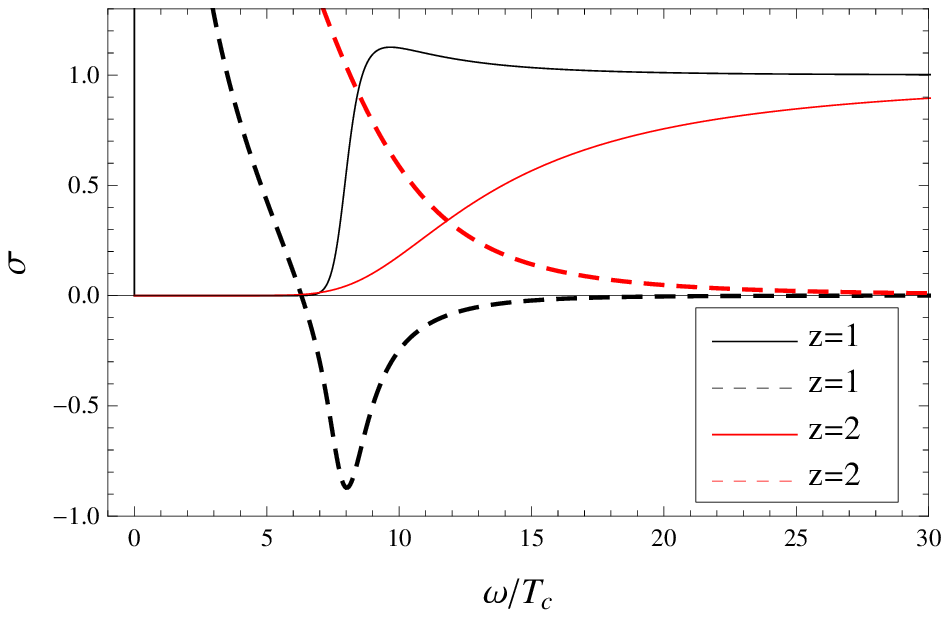}
\end{minipage}
\begin{minipage}[!htb]{0.43\linewidth}
\centering
\includegraphics[width=2.75in]{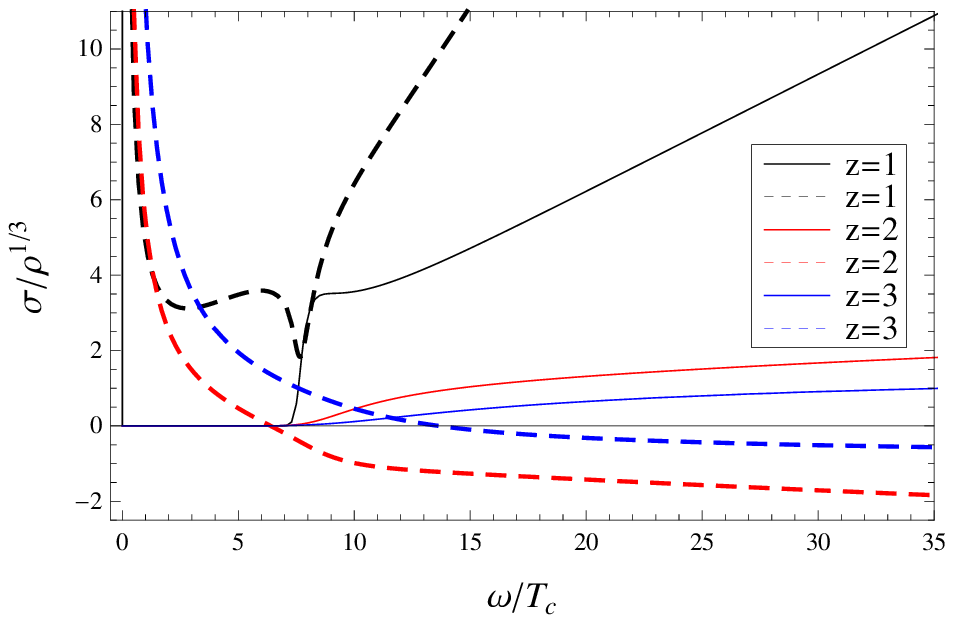}
\end{minipage}
\caption{The real (solid) and imaginary (dashed) part of the AC conductivity versus frequency of the $s$-wave model at $T/T_c\approx0.1$ with $\Delta=2$, $z=1$,~2 in $D=4$ (left), and $\Delta=3$, $z=1$,~2,~3 in $D=5$ (right).}
\label{d5zgdgm154bhSwaveCt}
\end{figure}

To compute the AC conductivity in the boundary field theory side, we need to study the perturbation of the gauge field in the bulk. Due to the rotational symmetry of the $s$-wave superconductor model, without loss of generality, we turn on the perturbation along the $x $ direction with the ansatz $\delta A_\mu=A_x(r)e^{-i \omega t}$. The linearized equation of the perturbation $A_x$ turns out to be
\begin{equation}\label{EomaxSwave}
 A_x''+\left(\frac{d+z-1}{r}+\frac{f'}{f}\right)A_x'+\frac{\omega^2}{r^{2z+2}f^2}A_x-\frac{2\psi^2}{r^2f}A_x=0.
\end{equation}
At the horizon, we impose the ingoing  wave condition
\begin{equation}
A_x(r)=(r-r_+)^{-i\omega/{4\pi T}}\left(1+A_{x1} (r-r_+)+A_{x2} (r-r_+)^2+A_{x3}(r-r_+)^3+\cdots\right).
\end{equation}
And at the boundary $r \to \infty$, the asymptotical expansion of $A_x(r)$ is of the form
\begin{equation}\label{bhSAxsolution}
A_x(r)=A^{(0)}+\frac{A^{(d+z-2)}}{r^{d+z-2}}+\cdots.
\end{equation}
Note that in the case of $z=1,D=5$, a logarithmic term $\frac{A^{(0)} \omega^2}{2r^2} \ln{\xi r}$ should be added to the right hand side of (\ref{bhSAxsolution}), where $\xi$ is a constant. According to the Kubo formula, the AC conductivity reads
\begin{eqnarray}\label{BHScondu}
\sigma(\omega)&=&\frac{-1}{i \omega}\lim_{r\rightarrow\infty}r^{d+z-1}\frac{A_x'}{A_x}.
\end{eqnarray}
In the case of $z=1$ in $D=5$, a logarithmic divergence exists in $\sigma(\omega)$, which can be canceled by the holographic renormalization~\cite{Horowitz126008}.
The AC conductivity is plotted in Fig.~\ref{d5zgdgm154bhSwaveCt}.  Here some remarks are in order.
 \begin{itemize}
 \item[(1)] There exists a pole in the imaginary part in $D=4~(5)$ at zero frequency. This pole corresponds to a delta function in the real part from the Kramers-Kronig relation, which is the signal of DC superconductivity.
 \item[(2)]In the case of $z=1$ in $D=4~(5)$, when the temperature decreases, there exists a sharp gap \footnote{In fact, this is an artifact of the probe approximation. Considering the backreaction from the matter field~\cite{Horowitz:2010gk,Horowitz:2009ij}, it was found that the real part of the conductivity at low frequency remains nonzero even at zero temperature, which does not satisfy the relation $Re[\sigma]\sim e^{-\Delta_g/T}$  with an energy gap $\Delta_g$, hence, there is no ``sharp gap" in the holographic superconductor.} frequency $\omega_g$. The ratio  $\omega_g/T_c \approx 8$,  much larger than the weak coupling BCS theory value 3.5,  indicates that the holographic model indeed describes a strongly coupled field theory. In addition, we can clearly see from Fig.~\ref{d5zgdgm154bhSwaveCt} that when $z \ne 1$ in $D=4$ and $D=5$, the minimum of the imaginary part of the conductivity disappears, which means that in those cases, the energy gap is no longer obvious.
 \item[(3)] In the case of $D=4$, we see from the left plot of Fig.~\ref{d5zgdgm154bhSwaveCt} that the real part of the conductivity is suppressed in the case of $z=2$, compared to the case of $z=1$. The same happens in the case of $D=5$, the conductivity for the case
     $ z>1$ is suppressed. This shows the anisotropic effect of the background spacetime.
  \item[(4)]In the $z=1$ and $D=5$ case,  both the real and imaginary parts of the conductivity diverge as $\omega \to \infty$, which is quit different from the corresponding case in $D=4$ dimensions. This behavior disappears when $z>1$. This is due to the absence of the logarithmic term in the expansion of $A_x$ near the boundary $r\to \infty$.
 \end{itemize}

\begin{figure}
\begin{minipage}[!htb]{0.45\linewidth}
\centering
\includegraphics[width=2.9in]{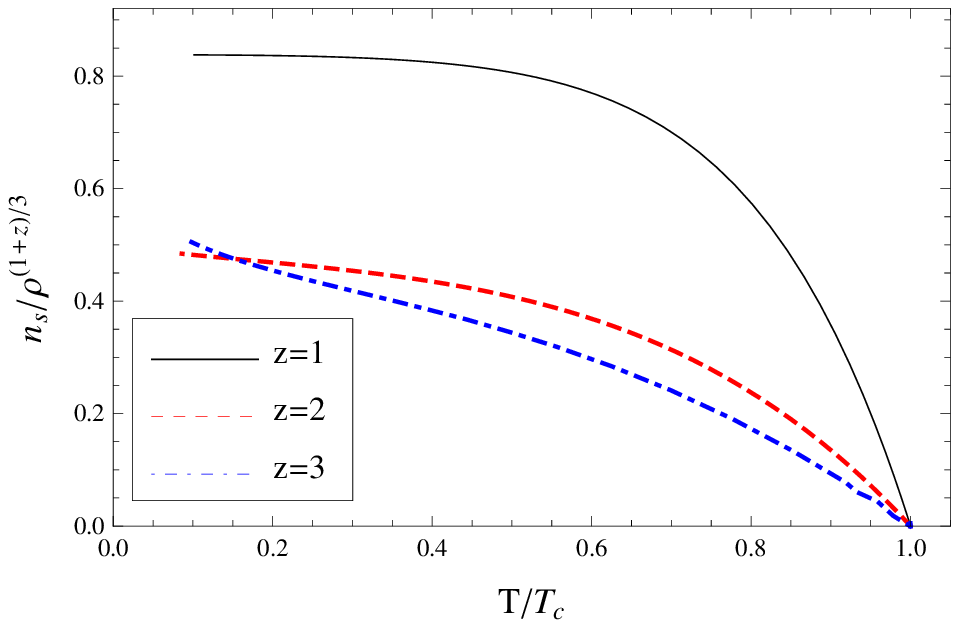}
\end{minipage}
\begin{minipage}[!htb]{0.45\linewidth}
\centering
\includegraphics[width=2.9in]{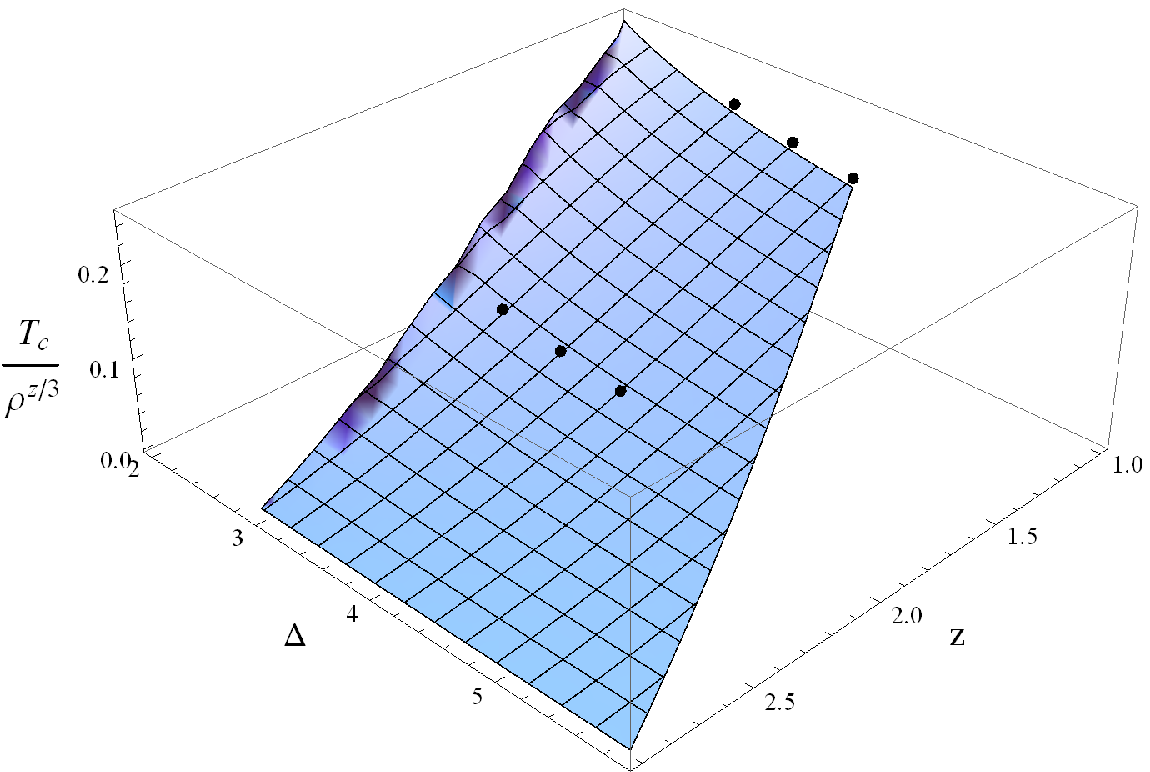}
\end{minipage}
\caption{The superfluid density versus temperature (left) with $\Delta=3$, $z=1,~2,~3$ (from top to bottom) and the critical temperature $T_c$ versus $z$ and $\Delta$ (right)  of the $s$-wave model in the 5-dimensional black hole background. The black dots in the right plot are from the numerical calculations with $z=1,~2$ for $\Delta=3,~7/2,~4$, respectively.}
\label{dgzgbhSanaTc}
\end{figure}

 The superfluid density $n_s$ can be calculated as the coefficient of the pole in $Im[\sigma]$ at $\omega=0$, i.e., $n_s\approx\lim_{\omega\rightarrow0}\omega Im[\sigma]$. In the left plot of Fig.~\ref{dgzgbhSanaTc}, we show $n_s$ with different $z$ and
 a fixed $\Delta=3$ in $D=5$ and we list $n_s$ near $T_c$ in Tab.~\ref{tab:BHswavenumericalresults} for all cases we calculated. We see that
 in all cases, $n_s$ has the behavior $n_s \sim (1-T/T_c)$ near the critical point. From the plot we see $n_s$ decreases with the increase of $z$. This is consistent with the result that the conductivity  decreases as $z$ increases, shown in Fig.~\ref{d5zgdgm154bhSwaveCt}.

Next we turn to the analytical study on the critical behavior of the $s$-wave superconductor model by employing the Sturm-Liouville variational  method~\cite{Siopsis}. Due to the presence of the logarithmic term in the falloff of the gauge field $\phi$ for the case of $z=d$, it is difficult to expand $\phi$ near the boundary. To avoid this, we here consider the case of $1\leq z<d$. Taking the new variable $u=r_+/r$, Eqs. (\ref{numbhspsi0inf}) and (\ref{numbhsphi0inf}) can be rewritten as
\begin{eqnarray}
\psi ''+\frac{u^{d+z}+d+z-1}{u \left(u^{d+z}-1\right)} \psi'+ \frac{m^2 \left(u^{d+z}-1\right)+r_+^{-2 z}u^{2 z} \phi^2}{u^2 \left(u^{d+z}-1\right)^2}\psi &=&0 \label{anabhspsi0inf},\\
\phi''+\frac{z-d+1}{u}\phi '+\frac{2 \psi ^2 }{u^2\left(u^{d+z}-1\right)}\phi&=&0,
\end{eqnarray}
where a prime represents the derivative with respect to $u$. As $T\rightarrow T_c$, the scalar field vanishes, so the solution of $\phi(u)$ is given by
\begin{equation}\label{anabhsphi01}
\phi(u)=\lambda r^z_{+c}\left(1-u^{d-z}\right),\lambda=\frac{\rho}{r^d_{+c}},
\end{equation}
where we have considered $\phi(1)=0$. We define a new function $F(u)$ as
\begin{equation} \label{anabhspsiasy}
\psi(u)=\frac{\langle\mathcal{O}\rangle}{r^{\Delta}_{+}}u^\Delta F(u).
\end{equation}
 We focus on the case of $D=5$. Substituting Eqs. (\ref{anabhsphi01}) and (\ref{anabhspsiasy}) into (\ref{anabhspsi0inf}), the latter can be
 rewritten as a Sturm-Liouville eigenvalue equation
\begin{equation}\label{eigenequationF}
\frac{d}{du}(\mathcal{T} F')-\mathcal{P}F +\lambda^2 \mathcal{Q}F=0,
\end{equation}
where  $\mathcal{T}, \mathcal{P}$, and $\mathcal{Q}$ read
\begin{eqnarray}
\mathcal{T}&=& \left(1-u^{z+3}\right) u^{\sqrt{4 m^2+z^2+6 z+9}+1},\nonumber\\
\mathcal{P}&=& \frac{1}{2} \left((z+3) \left(\sqrt{4 m^2+(z+3)^2}+z+3\right)+2 m^2\right) u^{\sqrt{4m^2+(z+3)^2}+z+2},\\
\mathcal{Q}&=& -\frac{u^{2 z}\left(u^3-u^z\right)^2 u^{\sqrt{4m^2+(z+3)^2}-2 z-1}}{u^{z+3}-1}.\nonumber
\end{eqnarray}
According to the boundary conditions for $F(u)$, i.e., $F(0)=1$ and $F'(0)=0$, we can introduce a trial function
\begin{equation} \label{trialfunction}
F=F_\alpha(u)\equiv 1-\alpha  u^2.
\end{equation}
The minimal eigenvalue $\lambda^2$ is obtained by minimizing the following expression with respect to the coefficient $\alpha$
\begin{equation}\label{eigenvalueexpress}
\lambda^2=\frac{\int_0^1du (\mathcal{T} {F'}^2-\mathcal{P}F^2)}{\int^1_0 du \mathcal{Q} F^2}.
\end{equation}
From the minimal eigenvalue of $\lambda^2$, we can read off the dependence of the critical temperature on the parameters $m^2$ and $z$. We show the analytical result on the critical temperature versus $z$ and $\Delta$ in the right plot of Fig.~\ref{dgzgbhSanaTc}.
 From the figure, we find the critical temperature decreases obviously when one increases $z$, while it decreases gradually with the increasing dimension of operator $\Delta$. Clearly the analytical result is consistent with the numerical one obtained above.  In Tab.~\ref{tab:BHswavenumericalresults} we list some critical temperatures $T_{c;SL}$ from the Sturm-Liouville variational method, to compare  with the numerical results.

When the temperature is slightly below $T_c$, the condensation $\frac{\langle\mathcal{O}\rangle^2}{r^{2\Delta}_{+}}$ is very small. In that case we can expand $\phi(u)$ in the form
\begin{equation}\label{bhsphichi}
\frac{\phi(u)}{r_+^z}=\lambda(1-u^{3-z})+\frac{\langle\mathcal{O}\rangle^2}{r^{2\Delta}_{+}}\chi(u)+\cdots.
\end{equation}
The equation of $\chi(u)$ reads
\begin{equation}
\chi ''+\frac{z-2}{u}\chi '-\frac{2 \lambda  \left(\alpha  u^2-1\right)^2 \left(u^3-u^z\right) u^{\sqrt{4
   m^2+(z+3)^2}+1}}{u^{z+3}-1}=0.
\end{equation}
Considering conditions $\chi(1)=0$ and $\chi'(1)=0$, we can have
\begin{equation}
\label{eq27}
u^{z-2}\chi(u)|_{u\rightarrow0}=2 \lambda \int^0_1 du \frac{\left(\alpha  u^2-1\right)^2 \left(u^3-u^z\right) u^{\sqrt{4m^2+(z+3)^2}+z-1}}{u^{z+3}-1}.
\end{equation}
Next expanding $\chi(u)$ near the boundary $u\rightarrow 0$, $\chi(u) =\chi(0)+ \chi'(0)u+\cdots$, we obtain
\begin{equation}
\frac{\rho}{r_+^{3}\lambda}-1=-\frac{\langle \mathcal{O} \rangle^2}{r_+^{2\Delta}}\frac{\chi^{3-z}(0)}{\lambda(3-z)!},
\end{equation}
by comparing the coefficient of $u^{3-z}$ in both sides of Eq. (\ref{bhsphichi}), where we have used  Eq.~(\ref{asySwaveBHphi}) and Eq.~(\ref{eq27}). Note that here $z$ is limited to be an integer,  $z=1$ or $2$.
Combining $r_{+c}=(\rho/\lambda)^{1/3}$ with the temperature (\ref{HawkingT}), we obtain
\begin{equation}\label{anabhcondent}
\langle\mathcal{O}\rangle^\frac{1}{\Delta}=\left(\frac{4\pi T_c}{3+z}\right)^\frac{1}{ z}\left(\frac{\lambda(3-z)!}{-\chi^{(3-z)}(0)}\right)^\frac{1}{2\Delta}
\left(1-(\frac{T}{T_c})^\frac{3}{z}\right)^\frac{1}{2\Delta}.
\end{equation}
We list the condensation in Tab.~\ref{tab:BHswavenumericalresults}, in order to compare with the numerical calculation. We see indeed
that the analytical calculation agrees with the numerical one at the same order. It might be worth stressing here that
the critical behavior of the condensation $\langle \mathcal{O}\rangle\sim(1-(T/T_c)^{d/z})^{1/2}$ looks a little different from the standard form $\sim(1-T/T_c)^{1/2}$. This is due to the scaling symmetry in the Lifshitz spacetime: $r\rightarrow \lambda r, T\rightarrow \lambda^z T, \langle \mathcal{O}\rangle\rightarrow \lambda^{\Delta} \langle \mathcal{O}\rangle, \rho\rightarrow\lambda^d\rho$. Further expanding (\ref{anabhcondent}) near the critical temperature, it is easy to see that the critical exponent is still $1/2$ and Eq.~(\ref{anabhcondent}) can be expressed as
\begin{equation}
\langle\mathcal{O}\rangle^\frac{1}{\Delta}=\left(\frac{3}{z}\right)^{\frac{1}{2\Delta}}\left(\frac{4\pi T_c}{3+z}\right)^\frac{1}{ z}\left(\frac{\lambda(3-z)!}{-\chi^{(3-z)}(0)}\right)^\frac{1}{2\Delta}
\left(1-\frac{T}{T_c}\right)^\frac{1}{2\Delta}.
\end{equation}

\subsection{$s$-wave superconductors in the Lifshitz soliton background}

In this subsection we consider the insulator/superconductor phase transition by generalizing the study in the AdS soliton
background~\cite{Nishioka131} to the Lifshitz soliton background with general $z$. By performing the double Wick rotation to the Lifshitz black hole solution (\ref{Lifshitz metric}), a $(d+2)$ dimensional Lifshitz soliton can be obtained as
\begin{equation}\label{Lifshitz soliton metric}
ds^2=-r^2dt^2+\frac{dr^2}{r^2f(r)}+r^2\sum_{i=1}^{d-1} dx_i^2+r^{2z}f(r)d\chi^2,
\end{equation}
where $f(r)$ still takes the form (\ref{metric function}). To distinguish the soliton from the black hole, we denote the tip of the soliton geometry by $r_0$.  To avoid a potential conical singularity at the tip, a periodicity on the spatial direction $\chi$ has to be imposed with a period $\chi\sim\chi+\frac{4\pi}{(z+d)r^z_0}$. For the soliton solution, there is no horizon, thus temperature is vanishing.  Due to the existence of the tip, there is an IR cutoff (mass gap) for the dual field theory. In other words, the dual field theory is in a confined phase. Thus similar to the case of the AdS soliton spacetime, the Lifshitz soliton solution can describe an insulator~\cite{Nishioka131}. In addition, let us notice that because of the compactness of the spatial direction $\chi$, this $(d+2)$-dimensional soliton geometry is dual to a $d$-dimensional field theory with mass gap, according to the gauge/gravity duality. In particular, we stress here that the Lifshitz soliton background (\ref{Lifshitz soliton metric}) does no longer have the anisotropic scaling $t \to b^z t$ and $\vec{x} \to b \vec{x}$ in the boundary spacetime as in the Lifshitz black hole background (\ref{Lifshitz metric}), but the dual boundary spacetime is of only the spatial anisotropy: $t\to bt$, $\vec{x}\to b\vec{x}$ and $\chi \to b^z \chi$.

Now we consider a holographic $s$-wave superconductor model based on the Lifshitz soliton background (\ref{Lifshitz soliton metric}).
The starting point of the matter sector is still the Lagrangian density (\ref{Swave action}). Due to the symmetry of the background, in the probe approximation, the ansatz for the matter sector is $A_t=\phi(r)$ and $\psi=\psi(r)$ following Ref.~\cite{Nishioka131}.  In the background (\ref{Lifshitz soliton metric}), the equations of motion for $\psi$ and $\phi$ read
\begin{eqnarray}
 \psi''+\left(\frac{d+z+1}{r}+\frac{f'}{f}\right)\psi'+\frac{\phi^2}{r^4f}\psi-
 \frac{m^2}{r^2f}\psi&=&0,\label{snumpsi}  \\
 \phi''+\left(\frac{d+z-1}{r}+\frac{f'}{f}\right)\phi'-\frac{2\psi^2}{r^2f}\phi &=& 0.\label{snumphi}
\end{eqnarray}
Interestingly, we see from the reduced equations of motion that the effective dimension of the spacetime increases from $(d+2)$ to $(d+z+1)$.
In other words, the equations of motion (\ref{snumpsi}) and (\ref{snumphi}) in the $(d+2)$-dimensional Lifshitz soliton background (\ref{Lifshitz soliton metric}) with
dynamical critical exponent $z$ are exactly the same as those in a $(d+z+1)$-dimensional AdS soliton background.
To solve the above two equations, we impose the Neumann-like boundary condition~\cite{Nishioka131} to make both $\psi(r_0)$ and $\phi(r_0)$ finite at the tip $r=r_0$. Near the boundary $r \to \infty$, $\psi(r)$ obeys the form (\ref{asySwaveBHSpsi}), while $\phi(r)$ is
\begin{equation}\label{asySPsolitionphi}
\phi(r)=\mu-\frac{\rho}{r^{d+z-2}}+\cdots,
\end{equation}
where the coefficients $\mu$ and $\rho$ are interpreted as the chemical potential and the charge density in the boundary field theory, respectively. We still take $r_0=1$ in the numerical calculation. Thus the period of the spatial coordinate $\chi$ is
 $\Gamma=4\pi/(z+d)$. To compare the effects of $z$ on the holographic superconductors, we rescale the period of $\chi$ to $\pi$.

\begin{figure}
\begin{minipage}[!htb]{0.45\linewidth}
\centering
\includegraphics[width=3.1in]{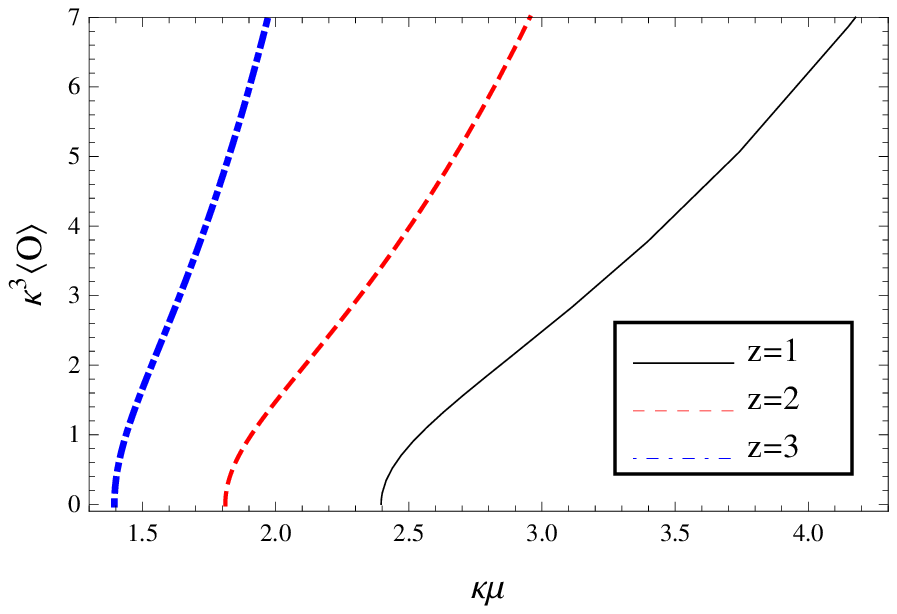}
\end{minipage}
\begin{minipage}[!htb]{0.45\linewidth}
\centering
\includegraphics[width=3.1in]{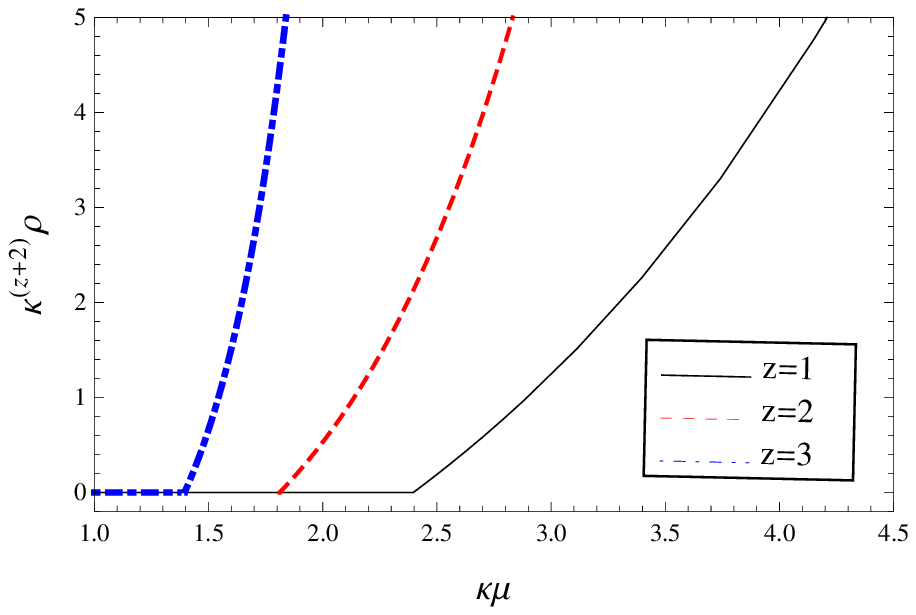}
\end{minipage}
\caption{The condensation (left) and the charge density (right) versus chemical potential of the $s$-wave superconductor with $\Delta=3$ and $z=1,~2,~3$ (from right to left) in the 5-dimensional Lifshitz soliton background, here $\kappa=(4/(3+z))^{1/z}$.}
\label{dgzgSScsCd}
\end{figure}

 From Eqs.~(\ref{snumpsi}) and (\ref{snumphi}), we see that this set of equations are determined by the parameters $(d+z)$ and $m^2$. Thus  the case of $z=1$ in $D=5$ is the same with the case of $z=2$ in $D=4$ if the mass of the scalar field is fixed. In Fig.~\ref{dgzgSScsCd}, we plot the condensation and charge density $\rho$ versus chemical potential in the case with $z=1,~2,~3$ and $\Delta=3$. We can see from the figure that when one fixes the dimension of the operator, the critical chemical potential decreases as $z$ increases. Near the critical point, we have $\langle \mathcal{O}\rangle \sim \sqrt{\mu-\mu_c}$ and $\rho\sim(\mu-\mu_c)$ by fitting the numerical curves in Fig.~\ref{dgzgSScsCd}. This means that the critical exponent of the condensation is still $1/2$, while the one for the charge density is one in all cases. This shows the universality of these critical exponents. In Tab.~\ref{tab:swavesolition} we list the critical chemical potential $\mu_c$ and condensation  in $D=4$ and $D=5$ for different mass of the scalar field, for a clear comparison.

\begin{figure}
\begin{minipage}[!htb]{0.45\linewidth}
\centering
\includegraphics[width=2.9 in]{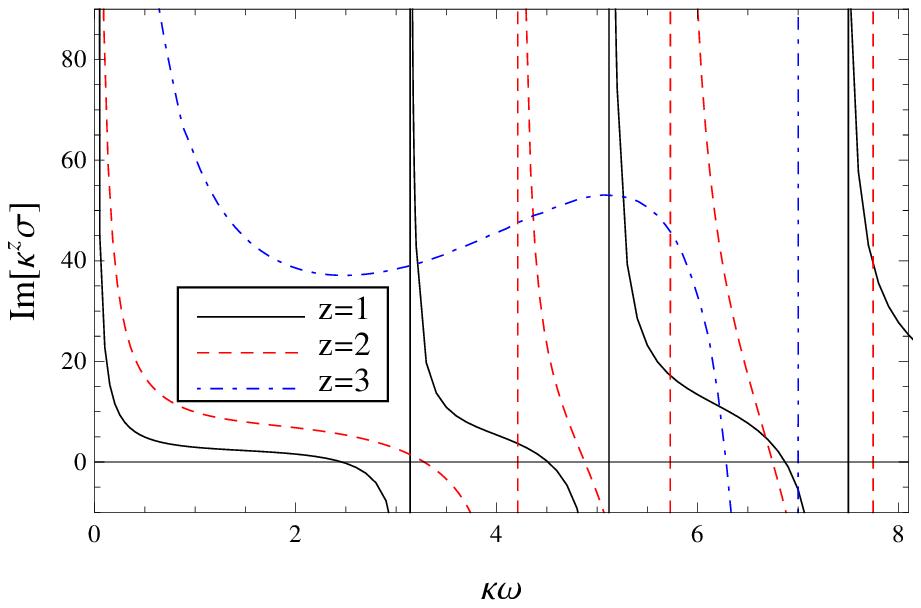}
\end{minipage}
\begin{minipage}[!htb]{0.45\linewidth}
\centering
\includegraphics[width=2.9 in]{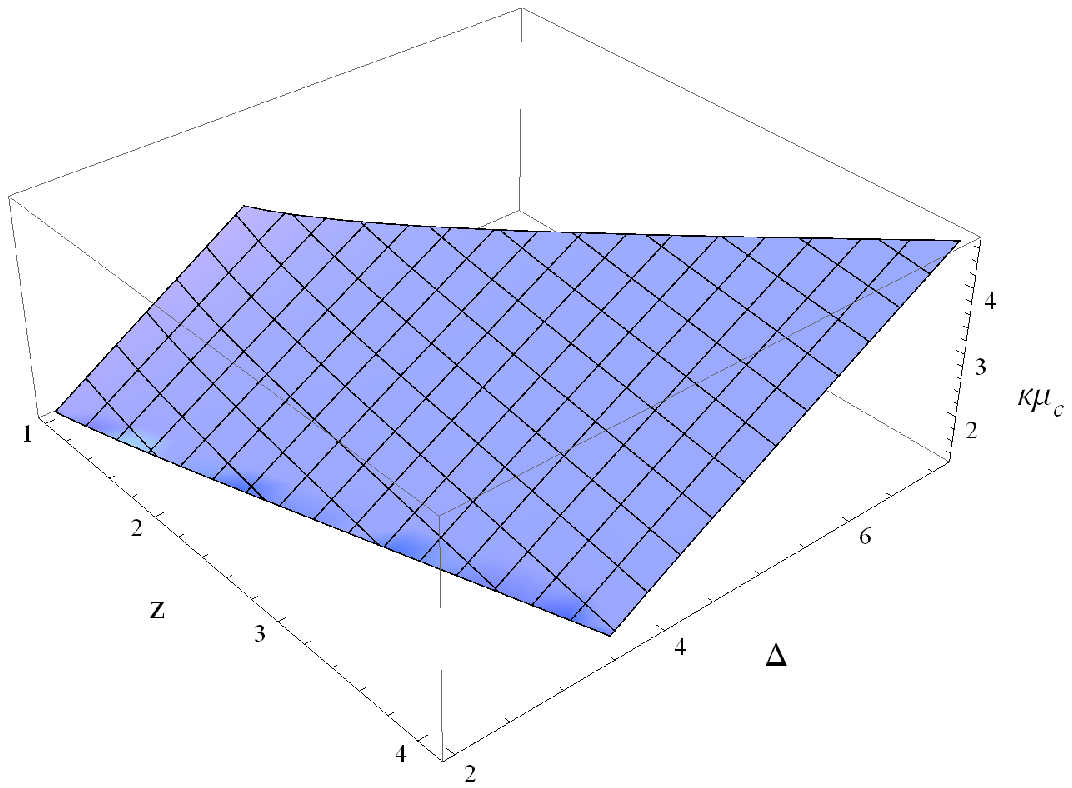}
\end{minipage}
\caption{The conductivity versus frequency with $\Delta=3$ and $z=1,~2,~3$ (from bottom to top in the most left) at $\mu/\mu_c\approx1.74$ (left) and the critical chemical potential $\mu_c$ (from the analytical calculation) as a function of $\Delta$ and $z$ (right) of the $s$-wave model in the 5-dimensional Lifshitz soliton background, here $\kappa=(4/(3+z))^{1/z}$. }
\label{dgzgmgsSct}
\end{figure}

 To calculate the conductivity, we turn on the perturbation $\delta A=A_x(r)e^{-i \omega t}$, which obeys the following equation
\begin{equation}\label{Sswaveaxeq}
 A_x''+\left(\frac{d+z-1}{r}+\frac{f'}{f}\right)A_x'+\frac{\omega^2}{r^4f}A_x-\frac{2\psi^2}{r^2f}A_x=0.
\end{equation}
In order for $A_x$ to be finite at the tip, we  take the ansatz of $A_x$ near the tip
\begin{equation}\label{Solitontipcond}
A_x(r)=1+A_{s1} (r-r_0)+A_{s2} (r-r_0)^2+A_{s3}(r-r_0)^3+\cdots.
\end{equation}
where $A_{s1}$, $A_{s2}$ and $A_{s3}$ are all constants and the leading term is taken to be unity due to the linearity of the equation for $A_x$.
Near the boundary $r\rightarrow\infty$, the general falloff of $A_x$ behaves as
\begin{equation}
\label{eq39}
A_x(r)=\left\{\begin{array}{ll}
  A^{(0)}+\frac{ A^{(1)}}{r}+\cdots, & z=1,~d=2,\\
   A^{(0)}+\frac{ A^{(2)}}{r^2}+\frac{ A^{(0)} \omega^2}{2r^2}\ln{\xi r}+\cdots, &z=2~(1),~d=2~(3),\\
   A^{(0)}+\frac{ A^{(0)}\omega^2}{2r^2}+\frac{ A^{(3)}}{r^3}+\cdots, &z=3~(2),~d=2~(3), \\
   A^{(0)}+\frac{ A^{(0)} \omega^2}{4r^2}+\frac{ A^{(4)}}{r^4}+\frac{ A^{(0)} \omega^4}{16r^4}\ln{\xi r}+\cdots, &z=3,~d=3,
  \end{array}
  \right.
  \end{equation}
where $A^{(i)}$, and $\xi$ are all constants. Using (\ref{BHScondu}), we  calculate the conductivity. Note that when the logarithmic term
appears in the expansion (\ref{eq39}), its effect on the conductivity can be removed by the holographic renormalization~\cite{Horowitz126008}, as in the previous subsection. We  plot the imaginary part of the conductivity in the case of $D=5$ in the left plot of Fig.~\ref{dgzgmgsSct}. The second pole positions in the imaginary part of the conductivity move toward the right as $z$ increases, which means
that the energy of the quasiparticle excitation increases as we increase $z$. The behavior of the conductivity in the $D=4$ case is similar as the one in the case of $D=5$. In Tab.~\ref{tab:swavesolition} we list the critical chemical potential, condensation, charge density and superfluid density in $D=4$ and $D=5$.
 We can see that the superfluid  density $n_s$ increases when $z$ increases, in the case with a fixed dimension $\Delta$ of the scalar operator. This indicates that in this case as $z$ increases the phase transition becomes easy. This is consistent with the observation that
 as $z$ increases the critical chemical potential decreases.
\begin{table}
\caption{\label{tab:swavesolition} The numerical and analytical results of the $s$-wave superconductors in the 4(5)-dimensional Lifshitz soliton backgrounds.  Here $t=(\mu/\mu_c-1)^{1/2}$, $\kappa=(4/(d+z))^{1/z}$, and the subscript ${SL}$ denotes the quantity calculated by the Sturm-Liouville variational method.}
\begin{ruledtabular}
\begin{tabular}{c c c c c   c c c}
  $D$& z & $m^2$ &$\kappa\mu_c$& $\kappa^\Delta\langle \mathcal{O}\rangle$  & $\kappa^{d+z-2}n_s$ & $\kappa\mu_{c;SL}$& $\kappa^\Delta\langle \mathcal{O}\rangle_{SL}$  \\ \hline
   4 &1& $0$ &  3.629& $5.32t$ & $1.50t^2$ &3.629&2.11$t$ \\
  4 &2& $-3$  &2.396 & $3.87t$ &$3.59t^2$ &2.399&2.82$t$ \\
   4 &3& $-6$  &1.880& $4.64t$ & $8.40t^2$ &1.884&3.54$t$ \\
  5 &1& $-15/4$  &1.888 & $3.39t$ &$8.48t^2$ &1.890&2.47$t$ \\
  5 &1& $-3$ & 2.396 & $3.85t$ & $3.56t^2$ &2.398&2.82$t$\\
  5&2& $-6$ &  1.811& $4.06t$ & $7.51t^2$ &1.815&3.48$t$\\
   5 &3& $-9$ &1.395 & $5.37t$ & $21.34t^2$ &1.400&4.13$t$ \\
\end{tabular}
\end{ruledtabular}
\end{table}

Now, we turn to the analytical calculation of the $s$-wave model in the $5$-dimensional Lifshitz soliton background. In the normal phase, the general solution $\phi(r)$ of Eq.~(\ref{snumphi}) is
\begin{equation}
\phi(u)=C_1 \left(\frac{1}{2 u^2}-\frac{\,_2F_1\left(1,-\frac{2}{d+z};1-\frac{2}{d+z};u^{d+z}\right)}{2u^2}\right)+C_2,
\end{equation}
where $C_1$ and $C_2$ are two constants, $_2F_1$ is the hypergeometric function and $u=r_0/r$. To have a regularity at the tip, we take
$C_1=0$ as in Refs.~\cite{Nishioka131, QYPan088, RGCai126007}. Thus $C_2=\mu$ giving the chemical potential in the dual field theory.
When $\mu$ is slightly beyond $\mu_c$, the condensation appears, the scalar field can be expressed as $\psi\approx \langle \mathcal{O}\rangle u^\Delta F(u)$, and  the function $F$ obeys the eigenvalue equation
\begin{equation} \label{SolitonSLeq}
\frac{d}{du}(\mathcal{T} F')-\mathcal{P}F +\mu_c^2 \mathcal{Q}F=0,
\end{equation}
where $\mathcal{T},\mathcal{P}$ and $\mathcal{Q}$ are given by
\begin{eqnarray}
\mathcal{T}&=&\left(1-u^{d+z}\right) u^{\sqrt{(d+z)^2+4 m^2}+1},\nonumber\\
\mathcal{P}&=&\frac{1}{2} \left((d+z)^2+(d+z)\sqrt{(d+z)^2+4 m^2}+2 m^2\right) u^{\sqrt{(d+z)^2+4m^2}+d+z-1},\\
\mathcal{Q}&=& u^{\sqrt{(d+z)^2+4 m^2}+1}.\nonumber
\end{eqnarray}
By taking the trial function (\ref{trialfunction}), the eigenvalue of $\mu_c^2$  is determined by minimizing the expression (\ref{eigenvalueexpress}). We show the critical chemical potential $\mu_c$ versus $\Delta$ and $z$ in the case of $D=5$ in the right plot of Fig.~\ref{dgzgmgsSct}.  We also calculate some cases in $D=4$ and $D=5$ as shown in Tab.~\ref{tab:swavesolition}. It can be seen that $\mu_c$ decreases with the increasing $z$, but it increases with the operator dimension $\Delta$. This supports our numerical calculations.

In the superconducting phase, near the critical point, the condensation $\langle \mathcal{O}\rangle^2$ is small, so we can expand $\phi(u)$ as
\begin{equation}\label{solitionphiansatz}
    \phi(u)=\mu_c+\langle \mathcal{O}\rangle^2 \chi(u)+\cdots.
\end{equation}
Note that the approximation $\psi\approx \langle \mathcal{O}\rangle u^\Delta F(u)$, we have  the equation of $\chi(u)$ as
\begin{equation}\label{sswavechizspecial}
\chi''+\frac{\left(3 u^{d+z}+d+z-3\right)}{u \left(u^{d+z}-1\right)} \chi'+\frac{2 \mu_c \left(\alpha u^2-1\right)^2 u^{\sqrt{(d+z)^2+4m^2}+d+z-2}}{u^{d+z}-1}=0.
\end{equation}
Defining  $T(u)=u^{-d-z+3} \left(1-u^{d+z}\right)$, Eq. (\ref{sswavechizspecial}) can be rewritten as
\begin{equation}
(T\chi')'-2 \mu_c \left(\alpha u^2-1\right)^2 u^{\sqrt{(d+z)^2+4 m^2}+1}=0.
\end{equation}
From the above equation, we can obtain
\begin{eqnarray}
\chi(0)&=&2 \mu_c\int_1^0\frac{dy}{T(y)}\int^y_1du \left(\alpha u^2-1\right)^2 u^{\sqrt{(d+z)^2+4 m^2}+1},\\
\chi^{(d+z-2)}(0)&=&2 \mu_c(d+z-3)!\int_1^0 du  \left(\alpha u^2-1\right)^2 u^{\sqrt{(d+z)^2+4 m^2}+1}.
\end{eqnarray}
Further, near the boundary $u\rightarrow 0$, $\phi(u)$ can be further expanded as~\cite{RGCai126007}
\begin{equation}\label{d5z2sscalarphiexp}
    \phi(u)= \mu_c+\langle \mathcal{O}\rangle^2\left(\chi(0)+\chi^\prime(0)u+\frac{1}{2}\chi^{\prime\prime}(0) u^2+\frac{1}{6}\chi^{\prime\prime\prime}(0)u^3+\cdots\right).
\end{equation}
Comparing the right hand side of Eq.~(\ref{d5z2sscalarphiexp}) with $\phi(u)= \mu- \rho u^{d+z-2}$, we obtain
\begin{eqnarray}
  \langle\mathcal{O}\rangle &=& \frac{1}{\sqrt{\chi(0)}}\sqrt{\mu-\mu_c},  \\
  \rho &=& \frac{\chi^{(d+z-2)}(0)}{(d+z-2)!\chi(0)}(\mu-\mu_c),
\end{eqnarray}
where we have used $\chi(1)=0$ as in Refs.~\cite{QYPan088, RGCai126007} and limited $z$ to be integer.  The condensation from the analytical method are listed in Tab.~\ref{tab:swavesolition}, which matches the numerical calculations at the same order.  When $m^2=-15/4$ and $z=1$ in $D=5$, the corresponding results recover the ones in Refs.~\cite{Nishioka131, RGCai126007}.

\section{Holographic $p$-wave superconductors in Lifshitz spacetime}

In this section, we first study the $p$-wave superconductor in the  Lifshitz black hole by numerical and analytical methods and then discuss the $p$-wave model in the Lifshitz soliton background. Following the proposals~\cite{Gubser2008a}, we construct the holographic $p$-wave superconductor model by considering an SU(2) gauge field in the bulk
\begin{equation}\label{Lpwave}
  \mathcal{L}_m=-\frac{1}{4}F^a_{\mu\nu}F^{a\mu\nu},
\end{equation}
where $F^a_{\mu\nu}$ is the field strength of the gauge field. The SU(2) group has three generators $\tau^i$ which satisfy the commutation relation $[\tau^i,\tau^j]=\epsilon^{ijk}\tau^k$ ($i,~j,~k=1,~2,~3$).   The equation of motion of the gauge field reads
\begin{equation}\label{EomofPwave}
   \nabla_\mu F^{a\mu\nu}+\epsilon^{abc}A^b_\mu F^{c\mu\nu}=0.
\end{equation}
In this model, a U(1) subgroup generated by $\tau^3$ is treated as the gauge group of electromagnetism, and  the gauge boson generated by $\tau^1$ charged by this U(1) is regarded as the vector field. The ansatz is~\cite{Gubser2008a}
\begin{equation}\label{BHPwaveansatz}
A=\phi(r) \tau^3 dt+\psi(r) \tau^1 dx.
\end{equation}
When $\psi \ne 0$, according to the gauge/gravity dictionary, the subleading term of this field at the boundary gives the vacuum expectation value of dual operator $J_x$.  The emergence of the non-trivial vector ``hair''  breaks the U(1) symmetry and the rotational symmetry, which mimics a $p$-wave superconducting phase transition in condensed matter physics.

\subsection{$p$-wave superconductors in the  Lifshitz black hole background}

As in the $s$-wave case, we work in the probe approximation. In this case, the equations of motion of $\phi(r)$ and $\psi(r)$  in  the Lifshitz black hole background (\ref{Lifshitz metric}) read
\begin{eqnarray}
 \psi''+\left(\frac{d+z-1}{r}+\frac{f'}{f}\right)\psi'+\frac{\phi^2}{r^{2z+2}f^2}\psi&=&0, \label{BHPwavepsi}  \\
 \phi''+\frac{d-z+1}{r}\phi'-\frac{\psi^2}{r^4f}\phi &=& 0.
\end{eqnarray}
For convenience in numerical calculations, we limit our consideration to the cases $1\leq z\leq d$ in $D=d+2=4$ and $5$ dimensions. At the black hole horizon, we require $\psi(r_+)$ to be regular and $\phi(r_+)=0$. Near the boundary $r \to \infty$, $\psi(r)$  behaves as
\begin{equation}\label{asyBHSPwavepsi}
\psi(r)=\psi_0+\frac{\langle J_x \rangle}{ r^{d+z-2}}+\cdots,
\end{equation}
while $\phi(r)$ takes the form (\ref{asySwaveBHphi}). We take $\psi_0=0$ since $\psi_0$ is regarded as the source term and the U(1) symmetry
is required to be broken spontaneously, while $\langle J_x \rangle $ is viewed as the vacuum expectation value of the vector operator with dimension $\Delta=d+z-1$ in the
boundary field theory.

\begin{figure}
\begin{minipage}[!htb]{0.45\linewidth}
\centering
\includegraphics[width=2.9in]{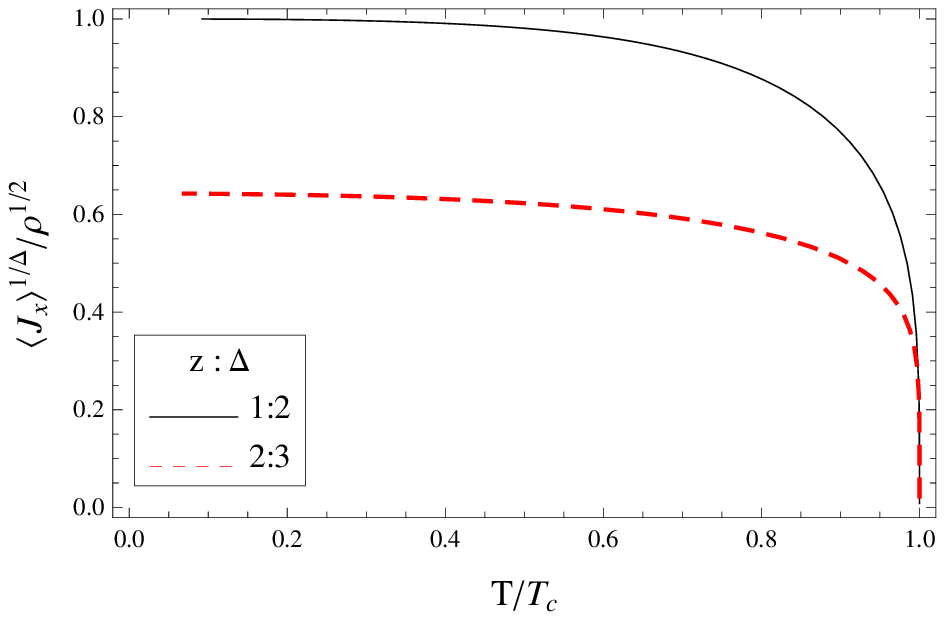}
\end{minipage}
\begin{minipage}[!htb]{0.45\linewidth}
\centering
\includegraphics[width=2.9in]{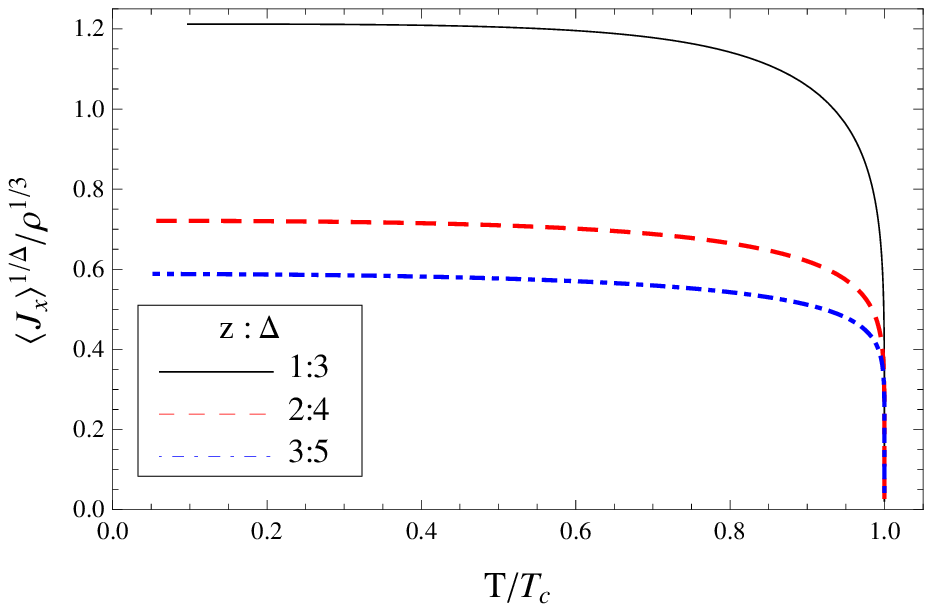}
\end{minipage}
\caption{The condensation  versus temperature for the $p$-wave model in $D=4$ (left) and  $D=5$ (right) Lifshitz black hole backgrounds.}
\label{d45zgbhpcondensation}
\end{figure}

We plot the  condensation in Fig.~\ref{d45zgbhpcondensation} and list some related results in Tab.~\ref{tab:BHPwavenumericalresults}.
We can see from the figure and the table that when $z$ increases, the critical temperature $T_c$ decreases, which implies that the increasing $z$ inhibits the phase transition. For a fixed temperature, the condensation decreases as $z$ increases, this means that the superconductivity becomes weak when we increase $z$. For all cases, near the critical point, the condensation behaves as $\langle J_x \rangle\sim(1-T/T_c)^{1/2}$. Once again, it shows the universality of the critical exponent.

\begin{figure}
\begin{minipage}[!htb]{0.45\linewidth}
\centering
\includegraphics[width=2.9in]{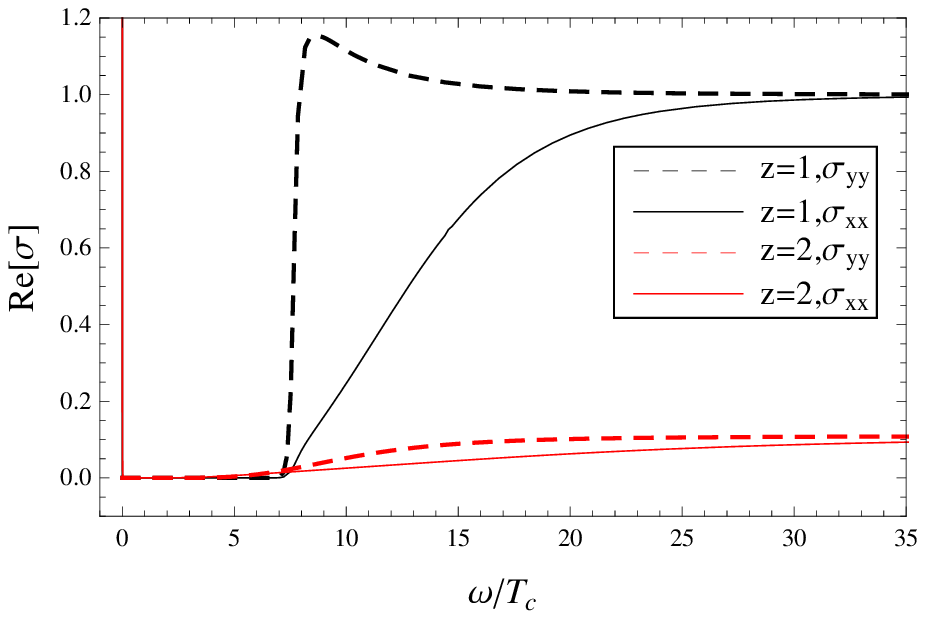}
\end{minipage}
\begin{minipage}[!htb]{0.45\linewidth}
\centering
\includegraphics[width=2.9in]{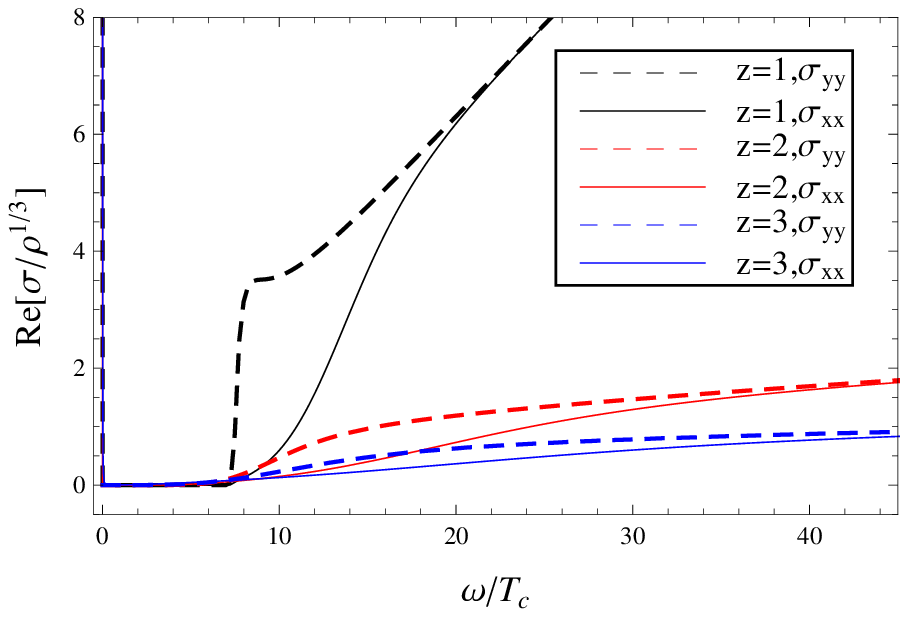}
\end{minipage}
\caption{The real part of the AC conductivity versus frequency of $a^3_y$ (dashed) and $a^3_x$ (solid) at $T/T_c\approx0.1$ in the black hole backgrounds. The left (right) plot corresponds to the case in $D=4~(D=5$) dimensions. }
\label{d45zgbhprect}
\end{figure}

\begin{figure}
\begin{minipage}[!htb]{0.45\linewidth}
\centering
\includegraphics[width=2.9in]{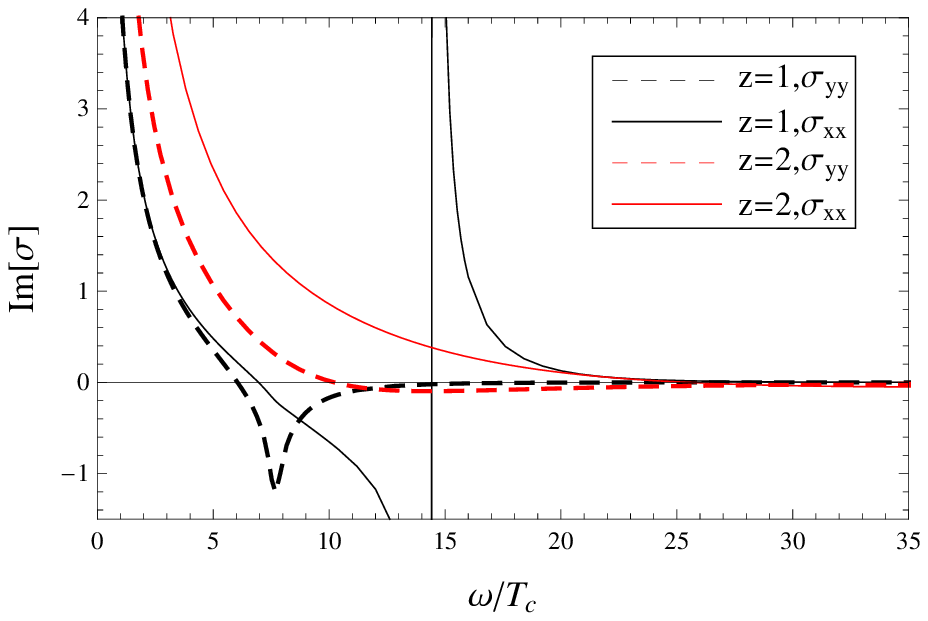}
\end{minipage}
\begin{minipage}[!htb]{0.45\linewidth}
\centering
\includegraphics[width=2.9in]{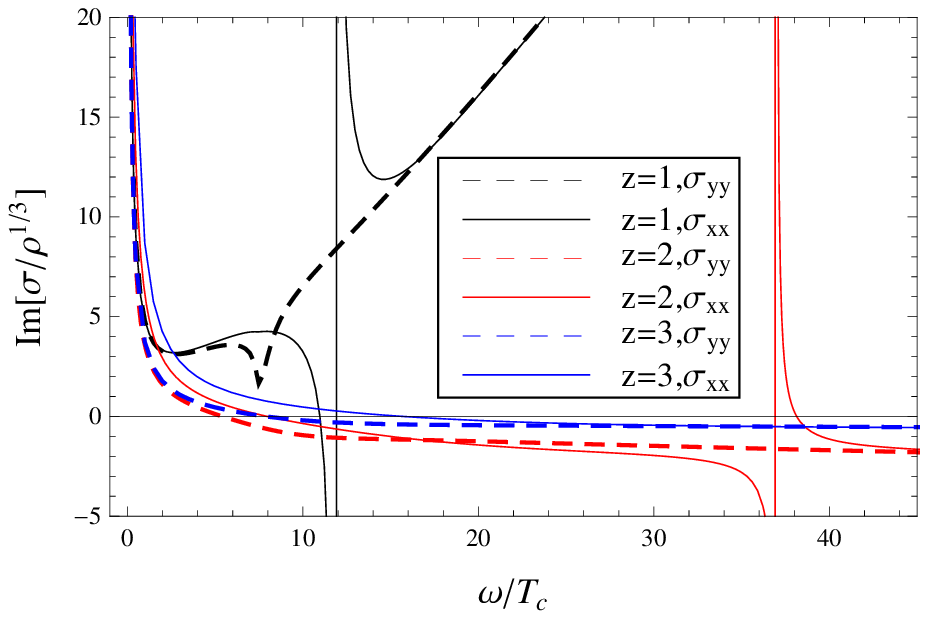}
\end{minipage}
\caption{The imaginary part of the AC conductivity versus frequency of $a^3_y$ (dashed) and $a^3_x$ (solid) at $T/T_c\approx0.1$ in the 4- (left) and 5-(right) dimensional black hole backgrounds with $z=1$,~2,~3.}
\label{d5zgbhpimct}
\end{figure}

The condensation of the vector field happens along the $x$ direction, so the conductivity $\sigma_{xx}$ along the $x$ direction  is expected to be different from that $\sigma_{yy}$ along the $y$ direction. To calculate the conductivity, we turn on the perturbation~\cite{Gubser2008a}
\begin{equation}\label{ansatzofPBHwave}
\delta A=e^{-i \omega t}\left(a^1_t(r) \tau^1 dt +a^2_t(r) \tau^2 dt+a^3_x(r) \tau^3 dx+a^3_y(r) \tau^3 dy\right),
\end{equation}
where we have taken the axial gauge $A^a_r=0$.
The linearized equations of motion of the Yang-Mills field result in four second order equations
\begin{eqnarray}
{a^3_y}^{\prime\prime}+\left(\frac{d+z-1}{r}+\frac{f'}{f}\right){a^3_y}'+\frac{\omega^2}{r^{2z+2}f^2}a^3_y-\frac{\psi^2}{r^4f}a^3_y &=&0,\label{EomofPwave}\\
  {a^1_t}''+\frac{d-z+1}{r}{a^1_t}'+\frac{\psi\phi}{r^{4}f}a^3_x&=& 0, \nonumber\\
  {a^2_t}''+\frac{d-z+1}{r}{a^2_t}'-\frac{\psi}{r^4f}(i\omega a^3_x+\psi  a^2_t) &=& 0, \label{EomofPaxx}\\
 {a^3_x}''+\left(\frac{d+z-1}{r}+\frac{f'}{f}\right){a^3_x}'-\frac{1}{r^{2z+2}f^2}(-\omega^2 a^3_x+i \omega \psi a^2_t+\psi \phi a^1_t)&=&0,\nonumber
\end{eqnarray}
as well as  two first order constraint equations
\begin{eqnarray}
  i\omega {a^1_t}'-\phi' a^2_t+\phi {a^2_t}' &=& 0, \\
  i \omega {a^2_t}'+\phi' a^1_t-\phi {a^1_t}'+r^{2z-2}f(\psi' a^3_x-\psi {a^3_x}') &=& 0,
\end{eqnarray}
where the prime denotes the derivative with respect to $r$. Obviously, $a^3_y$ is independent of other components, while $a^3_x$ mixes with $a^1_t$ and $a^2_t$. We impose the ingoing wave conditions at the horizon
\begin{equation}
a^3_y(r)=(r-r_+)^{-i\omega/{4\pi T}}\left(1+a_{y31} (r-r_+)+a_{y32} (r-r_+)^2+a_{y33}(r-r_+)^3+\cdots\right),
\end{equation}
\begin{eqnarray}
  a^1_t(r) &=&(r-r_+)^{-i\omega/{4\pi T}}\left(a_{t11} (r-r_+)+a_{t12} (r-r_+)^2+a_{t13}(r-r_+)^3+\cdots\right),\nonumber \\
  a^2_t(r) &=& (r-r_+)^{-i\omega/{4\pi T}}\left(a_{t21} (r-r_+)+a_{t22} (r-r_+)^2+a_{t23}(r-r_+)^3+\cdots\right),  \\
 a^3_x(r) &=& (r-r_+)^{-i\omega/{4\pi T}}\left(1+a_{x31} (r-r_+)+a_{x32} (r-r_+)^2+a_{x33}(r-r_+)^3+\cdots\right).\nonumber
\end{eqnarray}
On the other hand, near the boundary $ r \to \infty$, the expansion forms for these perturbations are listed in Tab.~\ref{tab:asybhp}.
\begin{table}
\caption{The asymptotical expansion of the perturbations of the $p$-wave model in the Lifshitz black hole background, where $i=1,~2$,  $\mu=x,~y$ and $\xi$ is a constant.}
\label{tab:asybhp}
\newcommand{\tabincell}[2]{\begin{tabular}{@{}#1@{}}#2\end{tabular}}
\begin{ruledtabular}
\begin{tabular}{c c c c c c c c }
   & $d=3,z=1$ & $d=3,z=2$ &$d=3,z=3$ & $d=2,z=1$ &$d=2,z=2$\\ \hline
  $a^i_t(r)$ & $a^{i(0)}_t+\frac{a^{i(2)}_t}{r^2}$ & $a^{i(0)}_t+\frac{a^{i(1)}_t}{r}$ & $a^{i(0)}_t+a^{i(1)}_t \ln{\xi r}$ & $a^{i(0)}_t+\frac{a^{i(1)}_t}{r}$ & $a^{i(0)}_t+a^{i(1)}_t\ln{\xi r}$  \\
  $ a^3_\mu(r)$ & \tabincell{c}{$a^{3(0)}_\mu+\frac{a^{3(2)}_\mu}{r^2}+$ \\$ \frac{a^{3(0)}_\mu\omega^2}{2r^2}\ln{\xi r}$ }& $a^{3(0)}_\mu+\frac{a^{3(3)}_\mu}{r^3}$ & $a^{3(0)}_\mu+\frac{a^{3(4)}_\mu}{r^4}$ & $a^{3(0)}_\mu+\frac{a^{3(1)}_\mu}{r}$ &  $a^{3(0)}_\mu+\frac{a^{3(2)}_\mu}{r^2}$  \\
  \end{tabular}
\end{ruledtabular}
\end{table}
As noticed in Ref.~\cite{Gubser2008a}, there still exists a gauge freedom in terms of $a^3_x$ . To calculate the conductivity along the $x$ direction, one can construct the gauge invariant quantity
 \begin{equation}\label{Pwaveax3hat}
 \hat{a}^3_x=a^3_x+\psi\frac{i \omega a^{2}_t+\phi a^{1}_t}{\phi^2-\omega^2}.
 \end{equation}
 And the conductivity can be expressed as
  \begin{equation}
 \sigma_{xx}(\omega)=\frac{\hat{a}^{3(1)}_x}{i \omega \hat{a}^{3(0)}_x},
 \end{equation}
where  $\hat{a}^{3(0)}_x$ and $\hat{a}^{3(1)}_x$ are the leading term and the coefficient of the sub-leading term of the expansion of $\hat{a}^3_x$ near the boundary.  The conductivity along the $y$ direction $\sigma_{yy}$ is of the same form as the one in (\ref{BHScondu}).
Note that when a logarithmic term appears in the conductivity, it is removed by the holographic renormalization as in the case of the $s$-wave model.  The conductivity $\sigma_{yy}$ and $\sigma_{xx}$ are plotted in Figs.~\ref{d45zgbhprect} and~\ref{d5zgbhpimct}.

Figure~\ref{d45zgbhprect} exhibits the real part of the conductivity. We can see that when $z=1$, there is an obvious energy gap from the conductivity in the $y$ direction, as the case in the $s$-wave model, while the energy gap becomes not so obvious in the $x$ direction. This is an obvious signature of the anisotropy for the $p$-wave superconductor model. Note that here the condensation appears in the $x$ direction. When $z$ increases, we can see clearly that in both cases with $D=4$ and $D=5$, the conductivity is obviously suppressed in both directions.
In those cases, the energy gap is also not very obvious. In addition, the difference between the conductivity along the $x$ and $y$ directions is also reduced as $z$ increases. This is the effect of the anisotropic scaling of the black hole background spacetime.

Figure~\ref{d5zgbhpimct} shows the imaginary part of the conductivity. We can see from the figure that when $z=1$ in $D=4$ and $D=5$, there exists an obvious minimum in the conductivity in the $y$ direction, while in other cases, the minimum disappears. This is consistent with the observation from the real part of the conductivity that when $z=1$, there exists an obvious energy gap along the $y$ direction, while it disappears in other cases. In addition, when $z=2$ in $D=4$ and $z=3$ in $D=5$  there does not exist a pole at a finite frequency in the imaginary part of the conductivity along the $x$ direction. The absence of the pole is due to the existence of the logarithmic term in the expansion of $\phi$ near the boundary. In fact, in this case the pole is pushed to the infinity of frequency, which can be see from (\ref{Pwaveax3hat}).

 We display the superfluid density $n_s^y~(n_s^x)$ along the $y~(x)$ direction in the left plot of Fig.~\ref{dgzgbhPanaTc}.
\begin{table}
\caption{ Some relevant quantities of
  the $p$-wave superconductor in the 4(5)-dimensional Lifshitz black hole backgrounds, where $t=1-T/T_c$. $\langle J_x\rangle^{1/\Delta}/\rho^{1/d}$, $\tilde{n}_s^y=n_s^y/\rho^{(d+z-2)/d}$ and $\tilde{n}_s^x=n_s^x/\rho^{(d+z-2)/d}$ are
calculated near $T_c$. The subscript ${SL}$ denotes the quantities calculated by the Sturm-Liouville variational method.}
\begin{ruledtabular}
\label{tab:BHPwavenumericalresults}
\begin{tabular}{c c  c c c  c c   c }
  $D$& z &$T_c/\rho^{z/d}$& $\langle J_x\rangle^{1/\Delta}/\rho^{1/d}$  &$\tilde{n}_s^y$ & $\tilde{n}_s^x$  & $T_{c;SL}/\rho^{z/d}$&$\langle J_x\rangle^{1/\Delta}_{SL}/\rho^{1/d}$\\ \hline
   4 &1 &$0.125$ & $1.40 t^{1/2\Delta}$ & $3.19 t$  &$1.04 t$  &0.124&$1.16 t^{1/2\Delta}$\\
    4 &2 &$0.037$ & $0.76 t^{1/2\Delta}$ & $0.93 t$ & $0.85 t$ &---&--- \\
5 &1 &$0.201$ & $1.62 t^{1/2\Delta}$ & $4.75 t$  & $2.37 t$ &0.199&$1.37 t^{1/2\Delta}$  \\
5 &2 &$0.065$ & $0.84t^{1/2\Delta}$ & $0.91 t$  & $0.90 t$ &0.065&$0.77 t^{1/2\Delta}$\\
5 &3 &$0.020$ & $0.65 t^{1/2\Delta}$ & $0.22 t$ & $0.36 t$  &---&---
\end{tabular}
\end{ruledtabular}
\end{table}
 We see that when $z$ increases, the superfluid density decreases. This indicates that the superconductivity becomes weak as $z$ increases, which is consist with the behavior of the condensation. Furthermore, near the critical point,  we see that
 $n_s^y~(n_s^x)$ behaves like $(1-T/T_c)$ in all cases.

Now we study the behavior of the $p$-wave model near the transition point by the Strum-Liouville variational  method. Near the critical point, the condensation is small, thus we can assume that the $\phi(u)$ obeys the form (\ref{anabhsphi01}), while the condensed field takes
\begin{equation}
\psi(u)=\frac{\langle J_x \rangle}{r_+^{d+z-2}} u^{d+z-2} F(u),
\end{equation}
where $F=1-\alpha u^2$. We can recast Eq.~(\ref{BHPwavepsi}) to the form of the eigenvalue problem equation (\ref{eigenequationF}) with
\begin{equation}
\mathcal{T}= u^{d+z-1} \left(1-u^{d+z}\right),\mathcal{P}= (d+z)(d+z-2) u^{2 d+2 z-3},
\mathcal{Q}= -\frac{ u^{d+z-3} \left(u^d-u^z\right)^2}{u^{d+z}-1}.
\end{equation}
As in the case of $s$-wave model, we can determine the critical temperature by the variational method. The critical temperature as a function of $z$ is plotted in the right plot of Fig.~\ref{dgzgbhPanaTc}. It follows that $T_c$ decreases when $z$ increases, which indicates that the increasing $z$ inhibits the phase transition. This is consistent with the numerical calculation.
\begin{figure}
\begin{minipage}[!htb]{0.45\linewidth}
\centering
\includegraphics[width=2.9in]{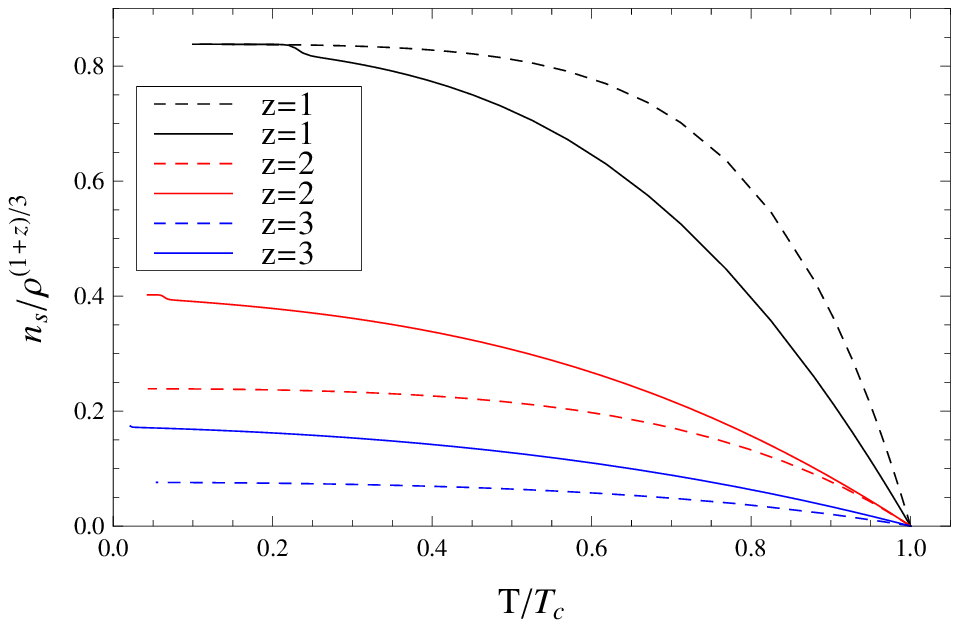}
\end{minipage}
\begin{minipage}[!htb]{0.45\linewidth}
\centering
\includegraphics[width=2.9in]{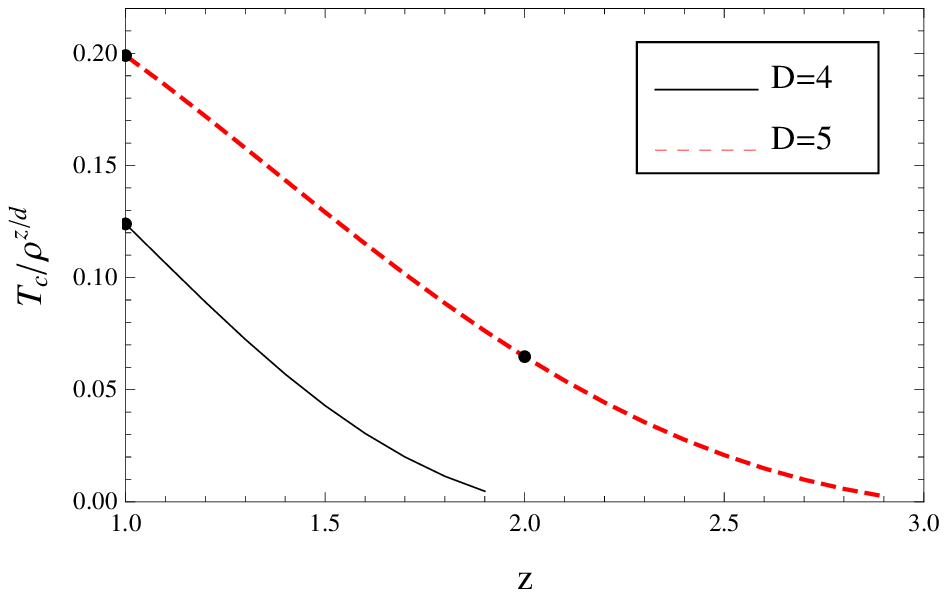}
\end{minipage}
\caption{The superfluid density versus temperature (left) and  the critical temperature $T_c$ versus $z$ (right) of the $p$-wave model. In the left plot, the solid curves denote $n_s^x$, while the dashed ones mean $n_s^y$.}
\label{dgzgbhPanaTc}
\label{dgzgbhPanaTc}
\end{figure}
In the same way, we can obtain the condensation behavior near the critical point. The critical temperature and condensation are listed in
Tab.~\ref{tab:BHPwavenumericalresults}. One can see that the analytical method gives consistent results with the numerical ones.

\subsection{$p$-wave superconductors in the Lifshitz soliton background}

\begin{figure}
\begin{minipage}[!htb]{0.45\linewidth}
\centering
\includegraphics[width=3.0in]{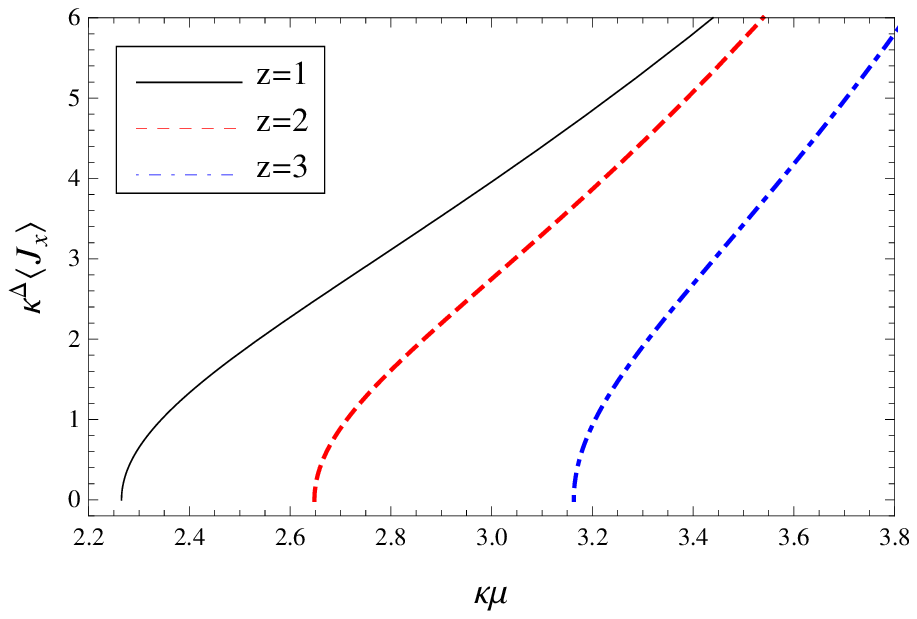}
\end{minipage}
\begin{minipage}[!htb]{0.45\linewidth}
\centering
\includegraphics[width=3.0in]{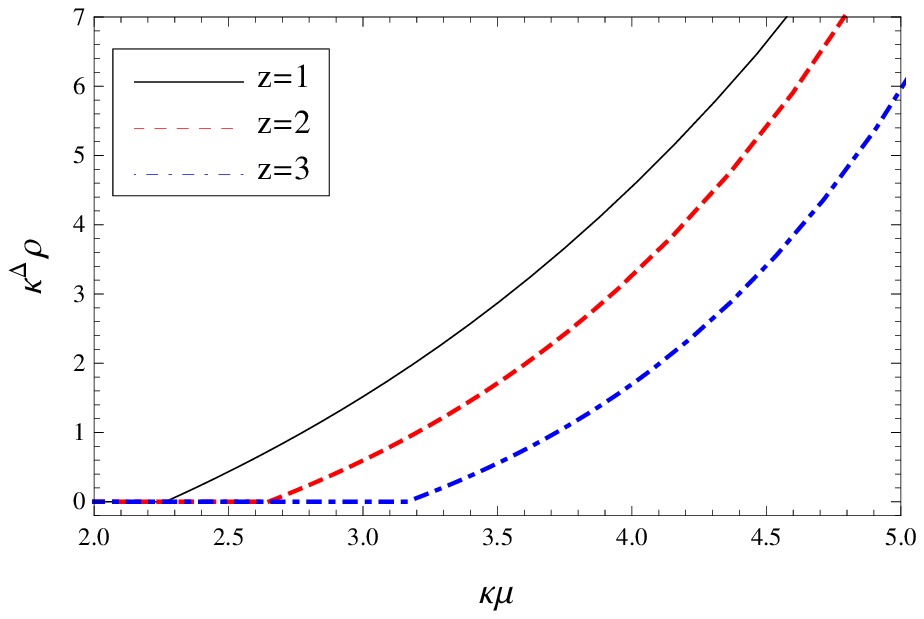}
\end{minipage}
\caption{The condensation (left) and the charge density (right) versus chemical potential of the $p$-wave model in a 5-dimensional soliton background, here $\kappa=(4/(3+z))^{1/z}$.}
\label{dgzgsPcscd}
\end{figure}

 In this subsection we consider a $p$-wave superconductor model in the Lifshitz soliton background (\ref{Lifshitz soliton metric}). This case
 corresponds to the insulator/superconductor phase transition at zero temperature. In the Lifshitz soliton background, the
 equations of motion of $\psi(r)$ and $\phi(r)$ turn out to be
\begin{eqnarray}
   \psi^{\prime\prime}+\left(\frac{d+z-1}{r}+\frac{f^\prime}{f}\right)\psi^\prime+\frac{\phi^2}{r^4f}\psi &=&0, \label{eomdgzgpwaveSpsi}\\
   \phi^{\prime\prime}+\left(\frac{d+z-1}{r}+\frac{f^\prime}{f}\right)\phi^\prime-\frac{\psi^2}{r^4f}\phi &=&0,\label{eomdgzgpwaveSphi}
\end{eqnarray}
It is interesting to note that these two equations are the same as those of the SU(2) Yang-Mills gauge field in a $(d+z+1)$-dimensional AdS soliton background.  This shows the equivalence between them in the probe approximation.
 Now we are going to solve these two equations. The boundary conditions of $\psi(r)$ and $\phi(r)$ at the tip $r=r_0$ are the same as those in the case of the $s$-wave model. On the other hand, near the boundary $ r \to \infty$, the general solution of $\psi(r)$  has the form (\ref{asyBHSPwavepsi}), while $\phi(r)$ takes the form (\ref{asySPsolitionphi}). We plot the
condensation $\langle J_x \rangle$  and the charge density $\rho$ in $D=5$  in Fig.~\ref{dgzgsPcscd}. It is easy to see that in this case, when $z$ increases, the critical chemical potential increases. This implies that when $z$ increases, the phase transition happens difficult.
 Near the critical point, we see that $\rho= C_\rho(\mu-\mu_c)$ and $\langle \mathcal{O}\rangle= C_\mathcal{O}\sqrt{\mu-\mu_c}$ in all cases. This shows, once again, the universal behavior of the critical exponent. In Tab.~\ref{tab:PwaveSnumerical} we list some results on the critical chemical potential and condensation with different $z$ in $D=4$ and $D=5$ cases.
  Note that when $z=1$ in $D=5$, our results can recover the ones in Refs.~\cite{Akhavan086003, RGCai046001, RGCai4828}.

In the presence of the condensation along the $x$ direction, the perturbation along the $x$ direction is expected to be mixed with other components. To calculate the conductivity along the $x$ direction $\sigma_{xx}$, we should turn on as many components as possible and then obtain a set of self-consistent equations for the perturbation. Considering the axial gauge $A_r^a=0$, we turn on the general perturbation in $D=5$ as
\begin{equation}\label{ansatzofPwave}
\delta A=e^{-i\omega t}a^a_\mu \tau^a dx^\mu,~a=1,~2,~3, \mu=t,~x,~y,~\chi.
\end{equation}
In the case of $D=4$, we have to turn off $a^a_y$ since this direction is absent in that case.
The linearized Yang-Mills equation results in the equation about $a^3_y$ (only in $D=5$)
\begin{equation}\label{Su2pSay3}
    {a^3_y}''+\left(\frac{f'}{f}+\frac{d+z-1}{r}\right){a^3_y}'+\frac{\omega ^2}{r^4 f}a^3_y-\frac{\psi^2 }{r^4 f}a^3_y=0,
\end{equation}
and three equations about $a^3_x$ (in $D= 4,5$)
\begin{eqnarray}
   {a^1_t}''+\left(\frac{f'}{f}+\frac{d+z-1}{r}\right){a^1_t}'+\frac{\psi\phi}{r^4f}a^3_x&=&0,\nonumber\\
{a^2_t}''+\left(\frac{f'}{f}+\frac{d+z-1}{r}\right){a^2_t}'-
\frac{\psi({a^2_t}\psi+i\omega {a^3_x})}{r^4f}&=&0,\\
   {a^3_x}''+\left(\frac{f'}{f}+\frac{d+z-1}{r}\right){a^3_x}'+\frac{\omega ^2 {a^3_x}-i \omega {a^2_t} \psi -{a^1_t} \psi  \phi }{r^4f}&=&0,\nonumber
 \end{eqnarray}
 as well as two constraint equations
 \begin{eqnarray}
  i\omega{a^1_t}'-\phi' a^2_t+\phi {a^2_t}' &=& 0, \\
  i \omega {a^2_t}'+\phi' a^1_t-\phi {a^1_t}'+\psi' a^3_x-\psi {a^3_x}' &=& 0.
\end{eqnarray}
 The other components decouple from $a^3_y$, $a^1_t$, $a^2_t$ and $a^3_x$. The boundary conditions of the above components at the tip are similar to those in (\ref{Solitontipcond}), while near the boundary $r \to \infty$, the general falloffs are listed in Tab.~\ref{tab:asySolitionP}.
\begin{table}
\caption{The asymptotical expansions of the perturbation of  $a^3_y$ and $a^3_x$ in 4~(5)-dimensional soliton background, where $\mu=x,~y$ in $d=3$ and $x$ in $d=2$, and $i=1, 2$.}
\label{tab:asySolitionP}
\newcommand{\tabincell}[2]{\begin{tabular}{@{}#1@{}}#2\end{tabular}}
\begin{ruledtabular}
\begin{tabular}{c c c c c c }
  & $d=3(2),z=1(2)$ & $d=3(2),z=2(3)$ &$d=3,z=3$ & $d=2,z=1$\\ \hline
  $a^i_t(r)$ & $a^{i(0)}_t+\frac{a^{i(2)}_t}{r^2}$ & $a^{i(0)}_t+\frac{a^{i(3)}_t}{r^3}$ & $a^{i(0)}_t+\frac{a^{i(4)}_t}{r^4}$ & $a^{i(0)}_t+\frac{a^{i(1)}_t}{r}$  \\
  $ a^3_\mu(r)$ & \tabincell{c}{$a^{3(0)}_\mu+\frac{a^{3(2)}_\mu}{r^2}$\\ $+\frac{a^{3(0)}_\mu\omega^2}{2r^2}\ln{\xi r}$} & $a^{3(0)}_\mu+\frac{a^{3(0)}_\mu \omega^2}{2r^2}+\frac{a^{3(3)}_\mu}{r^3}$ & \tabincell{c}{$a^{3(0)}_\mu+\frac{a^{3(0)}_\mu \omega^2}{4r^2}+\frac{a^{3(4)}_\mu}{r^4}$\\$+\frac{a^{3(0)}_\mu \omega^4}{16r^4}\ln{\xi r}$} & $a^{3(0)}_\mu+\frac{a^{3(1)}_\mu}{r}$  \\
  \end{tabular}
  \end{ruledtabular}
\end{table}

\begin{figure}
\begin{minipage}[!htb]{0.45\linewidth}
\centering
\includegraphics[width=2.9in]{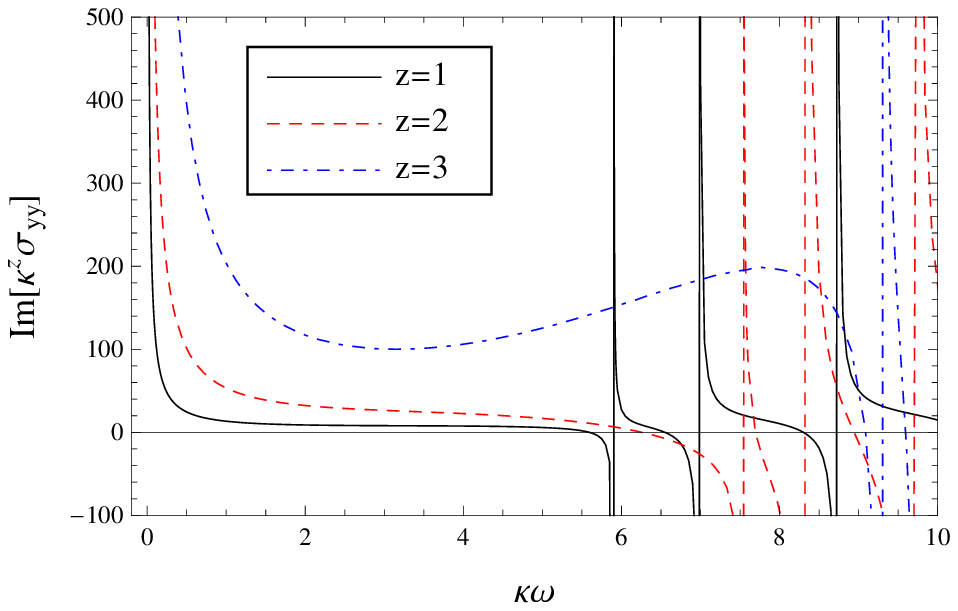}
\end{minipage}
\begin{minipage}[!htb]{0.45\linewidth}
\centering
\includegraphics[width=2.9in]{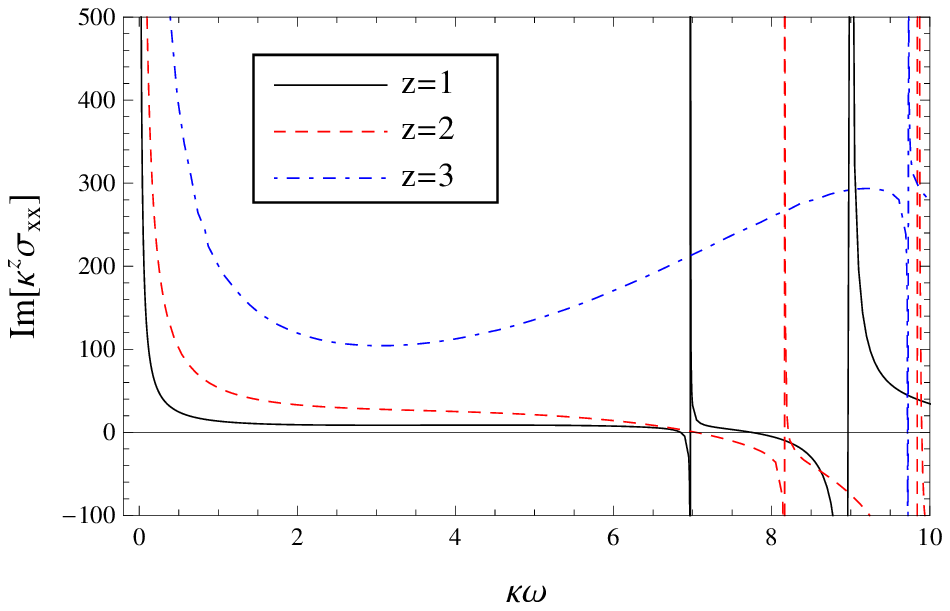}
\end{minipage}
\caption{The imaginary part of the  conductivity versus frequency of  $a^3_y$ (left) and $a^3_x$ (right) at $\mu/\mu_c\approx4$ in the 5-dimensional soliton background with $z=1,~2,~3$, here $\kappa=(4/(3+z))^{1/z}$.}
\label{d5zgsPaXXYYct}
\end{figure}

 In Fig.~\ref{d5zgsPaXXYYct} we plot the imaginary part of the conductivity along the $y$ direction (left plot) and the $x$ direction (right plot) for the $p$-wave model in $D=5$, while we list the superfluid density $n_s^y$~($n_s^x$) associated with $\sigma_{yy}~(\sigma_{xx})$  in Tab.~\ref{tab:PwaveSnumerical}, from which we have the following observations.
\begin{table}
\caption{\label{tab:PwaveSnumerical} Some quantities of the $p$-wave model in $D=4~(5)$, where $t=(\mu/\mu_c-1)^{1/2} $and quantities $\langle J_x\rangle$, $\langle J_x\rangle_{SL}$, $n_s^x$ and $n_s^y$ are all calculated near $\mu_c$. Here, $\kappa=(4/(d+z))^{1/z}$ and the subscript ${SL}$ denotes the quantity calculated by the Sturm-Liouville variational method.}
\begin{ruledtabular}
\begin{tabular}{c  c c  c  c c c  c}
  $D$& z &$\kappa\mu_c$& $\kappa^\Delta\langle J_x\rangle$ & $\kappa^{d+z-2}n_s^y$ & $\kappa^{d+z-2}n_s^x$ &$\kappa\mu_{c;SL}$& $\kappa^{\Delta}\langle J_x\rangle_{SL}$ \\ \hline
   4 &1  &1.988 & $4.12t$ &  -- &  $1.86t^{2}$ &1.988&2.24$t$ \\
    4 &2  &2.265 & $5.19t$  & -- & $3.54t^{2}$  &2.267&3.85$t$\\
  4 &3 &2.749 & $7.18t$ &  -- & $5.01t^{2}$ &2.752&$6.19t$ \\
 5 &1 &2.265 & $5.19 t$ &  $3.54t^{2}$&$3.54t^{2}$&2.267&$3.85t$ \\
  5 &2 & 2.648& $6.18t$ &  $4.07t^{2}$&$4.07t^{2}$& 2.652&$6.07t$\\
  5 &3 &3.163 & $8.38 t$ & $5.79t^{2}$ & $5.80t^{2}$ &3.171&9.12$t$\\
  \end{tabular}
\end{ruledtabular}
\end{table}
   Along the $y$ and $x$ directions, the position of the second pole in the imaginary part of the conductivity moves toward  the right as $z$ increases, which means that the energy of the quasiparticle excitation increases as we increase $z$. What is more, the frequency $\omega_s^x$ of the second pole in the $x$ direction is larger than $\omega_s^y$ in the $y$ direction in $D=5$. This indicates that to have a quasi-particle excitation along the $x$ direction (the condensing direction), one has to pay more energy than in the $y$ direction. In addition, we see that the difference between $\omega_s^x$ and $\omega_s^y$ decreases when $z$ increases, which indicates that the anisotropy of the conductivity is suppressed with the increasing $z$.

\begin{figure}
  \includegraphics[width=3.2 in]{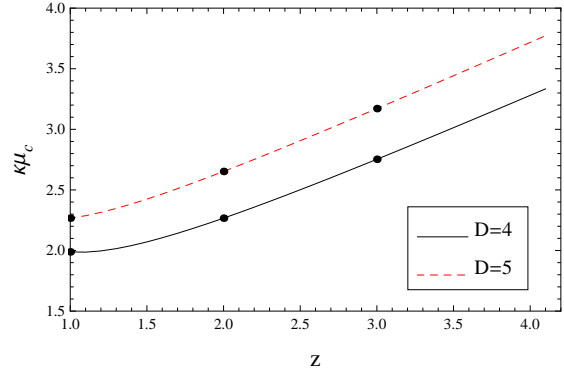}\\
  \caption{The critical chemical potential $\mu_c$ with respect to $z$  of the $p$-wave models in the Lifshitz soliton background, here $\kappa=(4/(d+z))^{1/z}$.}\label{dgzgSpmuc}
\end{figure}

 As in the $s$-wave model, we can also obtain a Sturm-Liouville eigenvalue equation (\ref{SolitonSLeq}) for the $p$-wave model in $D=5$ with
\begin{equation}\label{d5termpz}
   \mathcal{T}=u^{z+2} \left(1-u^{z+3}\right),\ \ \ \mathcal{P}=\left(z^2+4 z+3\right) u^{2 z+3},\ \ \ \mathcal{Q}=u^{z+2}.
\end{equation}
Solving Eq.~(\ref{SolitonSLeq}) with Eq.~(\ref{d5termpz}),  we can determine the critical chemical potential with arbitrary $z\geq1$, which is plotted in Fig.~\ref{dgzgSpmuc}.
The case of $D=4$ can be done in a similar way.
One can find that the critical chemical potential $\mu_c$ improves with the increase of $z$. This is consistent with the numerical calculation. Furthermore  we list the critical chemical potential and condensation in Tab.~\ref{tab:PwaveSnumerical}
from the analytical method, for a comparison with the numerical calculation. Once again, both methods show agreement results. This shows
that the analytical method is indeed powerful and universal. Finally we mention here that when $z=1$ and $D=5$, our results can recover
those  in Refs.~\cite{Akhavan086003, RGCai126007, HFLi135}.

\section{Conclusions and discussions}
In this paper, by numerical and analytical methods we have studied the properties of the holographic superconductor models in $4$- and $5$-dimensional Lifshitz black hole and soliton spacetimes, respectively. These models correspond to the conductor/superconductor and insulator/superconductor phase transitions in condensed matter physics. We have discussed the $s$-wave and $p$-wave models by introducing a complex scalar field and SU(2)
gauge field in the bulk, respectively. We have emphasized the influence of the dynamical critical exponent $z$ on the properties of the
holographic superconductor models. Main conclusions can be summarized as follows.
\begin{itemize}
\item[(1)] In the case with the Lifshitz black hole  background, for both the $s$-wave and $p$-wave models, when $z$ increases, the critical temperature decreases, which suggests that the phase transition becomes difficult as $z$ increases.  When $z$ increases, the conductivity
    is suppressed in both the $s$-wave and $p$-wave cases; the difference between the conductivity along the different directions, $\sigma_{yy}$ and $\sigma_{xx}$, decreases in the $p$-wave case. This indicates that when $z$ increases, the anisotropic effect in the $p$-wave model becomes weak. The superfluid density decreases as $z$ increases, which is consistent with the behavior of the conductivity.  But, we observed that near the
    critical point, the condensation always behaves as ~$\sim(1-T/T_c)^{1/2}$  in the case with a general $z$ and $D$. This shows
    that the critical exponent is universal, consistent with the result from the mean field theory.
\item[(2)] In the case with the Lifshitz soliton background, when $z$  increases, the critical chemical potential $\mu_c$ decreases in the $s$-wave models but increases in the $p$-wave cases. This result looks strange, but it can be understood by noting the fact that in the $s$-wave case, we fix the dimension of the scalar operator, while in the $p$-wave case, the mass of the vector field is fixed (in fact, the effective mass of the vector field is zero).  In Fig.~\ref{dgzgSpmucfixedm} we plot the critical chemical potential $\mu_c$ with various $z$ and mass the scalar field in the $s$-wave model. We can see clearly that for a fixed mass, the critical chemical potential
    increases when $z$ increases, which is in agreement with the case of the $p$-wave model. In the $p$-wave model, we have found that the difference between $\sigma_{xx}$ and $\sigma_{yy}$ decreases with the increase of $z$ , which implies that the increasing $z$ suppresses the anisotropy of the $p$-wave superconductor. In addition, near the critical point, the condensation always behaves
    as $\sim (\mu-\mu_c)^{1/2}$ in all cases. Once again, this shows the universality of the critical exponent.
 \begin{figure}
  \includegraphics[width=3.2 in]{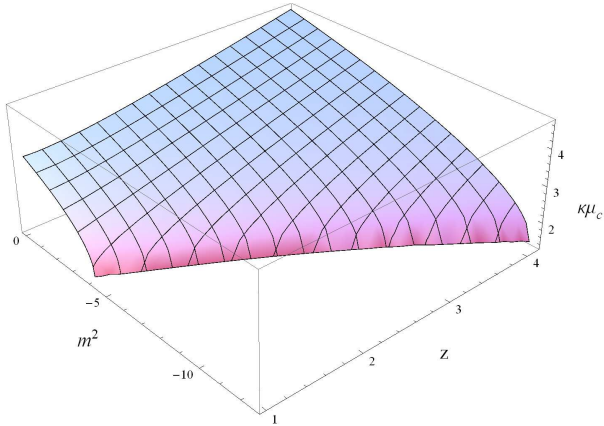}\\
  \caption{The critical chemical potential $\mu_c$ versus $z$ and $m^2$ of the $s$-wave models in the 5-dimensional soliton.}\label{dgzgSpmucfixedm}
\end{figure}
\item[(3)] In both the cases with the black hole and soliton backgrounds, by employing the Sturm-Liouville variational method, we have
studied the behavior of these holographic superconductor models near the critical point, and obtained consistent results as in the numerical calculation. This shows that the variational method is quite useful and powerful.
 \end{itemize}

In a word, we have seen that when the dynamical critical exponent $z$ increases, the superconducting phase transition becomes difficult,
 the superconductivity becomes weak and in the $p$-wave case the anisotropy is suppressed. In addition, we have found that in the $(d+2)$-dimensional Lifshitz soliton background, the reduced equations of motion in both the $s$-wave and $p$-wave models are the same as
 those in a $(d+z+1$)-dimensional AdS soliton background. As a result, the superconductor models with the $(d+2)$-dimensional Lifshitz soliton background are equivalent to those in a $(d+z+1$)-dimensional AdS soliton background. Of course, this holds only in the probe approximation.

In this paper we have only worked on the probe limit by neglecting the backreaction of the matter fields. Although the probe limit can reveal some significant properties of holographic superconductor model, it has been shown that new phases can emerge (see Refs.~\cite{Cai:2013wma,Liu:2013yaa} for example) and the order of the phase transition can also be changed~\cite{Horowitz:2010jq,Cai:2012es} once the backreaction is taken into consideration. Therefore, it is interesting to study the influence of the backreaction of the matter field to the Lifshitz background and to see whether there are some new features beyond the probe limit.

\acknowledgments  We would like to thank S.~A.~Hartnoll for his help about the numerical code.  One of the authors (J.~W.~Lu) is  deeply grateful to Y.~Y.~Bu, L.~Li,  and M.~L.~Liu  for their helpful discussions and comments, especially to R.~G. Cai for his directive help. This work is supported by the National Natural Science Foundation of China (Grant No.~11175077), the Joint Specialized Research Fund for the Doctoral Program of Higher Education, Ministry of Education, China (Grant No.~20122136110002), the Project of Key Discipline of Theoretical Physics of Department of Education in Liaoning Province (Grant Nos.~905035 and 905061), and the Open Project Program of State Key Laboratory of Theoretical Physics, Institute of Theoretical Physics, Chinese Academy of Sciences, China (No.~Y4KF101CJ1).

\end{document}